%% file: main.tex
\documentclass[aps,prd,twocolumn,10pt,superscriptaddress,amssymb,amsmath,nofootinbib]{revtex4-2}
\usepackage[utf8]{inputenc}
\usepackage{lmodern}
\usepackage[T1]{fontenc}
\usepackage{graphicx} 
\usepackage[breaklinks,colorlinks,
linkcolor=Blue,
citecolor=teal,
anchorcolor=red,
urlcolor=cyan,
pdfencoding=auto]{hyperref}
\usepackage[dvipsnames]{xcolor}
\usepackage{mathtools}
\usepackage{mathrsfs}
\usepackage{booktabs}
\usepackage{longtable}
\usepackage{soul}
\usepackage{orcidlink}
\usepackage[mathscr]{euscript}
\newcommand{\beq}{\begin{equation}\begin{aligned}}
\newcommand{\eeq}{\end{aligned}\end{equation}}

\def\ccr#1{\textcolor{red}{#1}}

\hypersetup{colorlinks=true}

\newcommand{\rhosp}{\rho_{\mathrm{sp}}}
\newcommand{\rsp}{r_{\mathrm{sp}}}

\newcommand{\avdot}[1]{\left\langle\frac{d#1}{dt}\right\rangle}

\newcommand{\PNx}{x}

\newcommand{\kai}{\chi}

\newcommand{\IDMzero}{\mathcal{I}_{0}}
\newcommand{\IDMone}{\mathcal{I}_{1}}

\def\equationautorefname~#1\null{%
	Eq.~(#1)\null
}
\def\figureautorefname~#1\null{%
	Fig.~#1\null
}
\def\tableautorefname~#1\null{%
	Table.~#1\null
}
\def\sectionautorefname~#1\null{%
	Section #1\null
}
\def\appendixautorefname~#1\null{%
	Appendix #1\null
}

\begin{document}
\raggedbottom

\title{Constraint on environments with eccentric extreme-mass-ratio inspirals: Bayesian inference and Fisher-matrix for dark-matter spikes}

\author{Genliang Li\,\orcidlink{0009-0008-9893-6160}}
\affiliation{College of Computer and Big data, Putian University, Putian 351100, Fujian, China}

\author{Tieguang Zi\,\orcidlink{0000-0003-0046-2056}}
\affiliation{Department of physics, Nanchang University, Nanchang 330031, Jiangxi, China}
\affiliation{Center for Relativistic Astrophysics and High Energy Physics, Nanchang University, Nanchang, 330031, China}

\date{\today}

\begin{abstract}
Extreme-mass-ratio inspirals (EMRIs) provide sensitive probes of the
astrophysical environments surrounding massive black holes. We investigate
the inference of dark-matter (DM) spikes with eccentric EMRI gravitational
waves within the analytic-kludge waveform framework, including
DM-induced dynamical friction and accretion in the secular orbital
evolution. We perform Bayesian parameter estimation using Markov-chain
Monte Carlo sampling, complemented by Fisher-matrix analyses. We find that
steeper DM spikes produce stronger accumulated waveform modifications and
substantially improve the measurement of the spike slope
\(\alpha_{\rm DM}\). More importantly, the constraints exhibit a
nonmonotonic dependence on orbital eccentricity. At moderate eccentricity,
higher harmonics help break the degeneracy between \(e_{\rm LSO}\) and
\(\alpha_{\rm DM}\), whereas this correlation can persist or the parameter
recovery can deteriorate for highly eccentric systems. These results
demonstrate that the harmonic structure of eccentric EMRIs can provide
valuable information for probing DM distributions around massive black
holes with future LISA observations.
\end{abstract}

\maketitle

\section{Introduction}\label{Introduction}
The rapidly growing catalog of compact-binary coalescences observed by
the LIGO--Virgo--KAGRA (LVK) Collaboration has established gravitational
waves (GW) as a new probe not only of the compact objects themselves, but also
of the astrophysical environments in which they form and merge
\cite{Barausse:2014tra,Cardoso:2019rou,Toubiana:2020drf,Zwick:2021dlg,KAGRA:2021vkt,Sberna:2022qbn,CanevaSantoro:2023aol,Roy:2024rhe,DuttaRoy:2025gnu,Duque:2025yfm,HegadeKR:2025dur,HegadeKR:2025rpr}. Although standard GW analyses generally assume isolated binaries evolving in vacuum, compact binaries may reside in dense stellar systems, active-galactic-nucleus disks, gaseous media, or regions containing dark-matter (DM) overdensities and ultralight bosonic fields \cite{Barausse:2014tra,Barausse:2014pra,Cardoso:2019rou,Duque:2023seg}. Interactions with the surrounding medium, including dynamic friction (DF), mass accretion, migration torques, and the gravitational potential of the ambient matter, can modify the binary dynamics and imprint additional phase and amplitude corrections on the observed waveform~\cite{Yunes:2011ws,Cardoso:2019rou}.
Recent analyses of LVK observations have begun to search directly for such effects, placing bounds on the density of matter surrounding compact binaries, on scalar-field configurations around binary black holes, and on nearby compact perturbers~\cite{CanevaSantoro:2023aol,Roy:2025qaa,Giri:2026wgy}. No compelling evidence for environmental effects has yet been established, but these studies demonstrate that GW observations can be used to constrain nonvacuum environments surrounding black holes. Dense DM spikes are of particular interest to GW astronomy. DF, accretion, and the additional gravitational potential generated by a spike can modify the inspiral rate and produce an accumulated GW dephasing~\cite{Eda:2013gg,Eda:2014kra,Kavanagh:2020cfn,Coogan:2021uqv,Karydas:2024fcn}. These effects are especially important for intermediate- and extreme-mass-ratio inspirals, whose long observation times and large numbers of orbital cycles allow otherwise weak environmental corrections to
accumulate to potentially measurable levels~\cite{Li2022,Cole:2022ucw,Zi:2025onl,Zi:2026zpw}.

Extreme-mass-ratio inspirals (EMRIs), in which a stellar-mass compact
object inspirals into a massive black hole (BH) with a characteristic
mass ratio of \(q\sim10^{-4}-10^{-7}\), are among the most important
targets of future space-based GW observatories,
including LISA~\cite{Babak_2017}, Taiji~\cite{Hu2017}, and
TianQin~\cite{Luo_2016}. During their evolution through the millihertz
frequency band, EMRIs can accumulate
\(\mathcal{O}(10^{4}-10^{5})\) orbital cycles before plunge
\cite{amaro2007intermediate,Babak_2017}. Consequently, even weak
perturbations to the orbital dynamics can accumulate over long
observation times and leave measurable imprints on the GW phase.
EMRI signals therefore encode detailed information about the
central spacetime, the intrinsic source parameters, and the
astrophysical environment surrounding the massive BH, making these
systems powerful probes of strong-field gravity, black-hole physics,
and the astrophysical environment surrounding massive BHs
\cite{cardoso2019testing,amaro2007intermediate,Babak_2017,Tahelyani:2024cvk,Li:2025zgo,Machet:2026ozy}.

Among possible environmental effects, dark matter (DM) in the vicinity of
massive BHs is of particular interest. The existence of DM is the most important candidate for the current astrophysical and cosmological observations
\cite{bertone2005particle,bertone2018history}. If a massive BH grows
adiabatically within a preexisting DM halo, the surrounding density can be
strongly enhanced, leading to the formation of a DM spike
\cite{Gondolo1999}. Whereas conventional DM searches mainly probe
nongravitational interactions or large-scale DM properties, GW observations
provide a complementary way to investigate DM distributions in the immediate
vicinity of BHs~\cite{cardoso2019testing}.

A compact object inspiralling through a DM spike is affected by dynamical
friction (DF), accretion, and modifications of the background gravitational
potential. These effects can alter the inspiral rate and generate accumulated
phase shifts in the emitted GW signal
\cite{Macedo_2013,Eda:2013gg,Eda:2014kra}. Subsequent studies have shown that the
dynamical response of the DM distribution can further modify the waveform
phase and the detectability of the environmental imprint
\cite{Yue_Han_2018,Kavanagh:2020cfn,Coogan:2021uqv,Vicente:2025gsg,Duque:2023seg}. For eccentric inspirals, DF can significantly affect
the eccentricity evolution and produce waveform differences relative to the
vacuum case~\cite{Yue_Han_Chen_2019,Zwick:2025ine}.
More recent developments have also incorporated the dynamical response of the
DM distribution, accretion feedback, improved DF prescriptions, and
relativistic treatments of collisionless environments
\cite{Dosopoulou_2024,Karydas:2024fcn,Vicente:2025gsg,Li:2025qtb}.

Many existing parameter-estimation studies have adopted circular or
quasicircular orbital models and have mainly relied on waveform dephasing,
mismatch calculations, or Fisher-information-matrix (FIM) forecasts~\cite{Eda:2014kra,Rahman:2023sof,Zhang_2024,Gliorio_2025}.
Astrophysically formed EMRIs, however, can retain non-negligible eccentricity in the observational band of space-based
detectors~\cite{BarackCutler2004,Babak_2017,Mancieri_2025}. Eccentricity
introduces additional harmonics and modifies both the amplitude and phase
evolution of the waveform, potentially affecting the degeneracy between the
DM parameters and the intrinsic orbital parameters. The combined effects of orbital eccentricity and parameter degeneracies in DM-dressed EMRIs remain comparatively less explored.

This motivates a Bayesian parameter-estimation study of DM effects in
eccentric EMRIs. Although the FIM provides a useful local estimate of
parameter uncertainties, it cannot fully capture non-Gaussian posterior
structures or nonlinear parameter degeneracies. Existing studies of eccentric
IMRI dynamics and waveforms in DM minispikes provide an important foundation
for this direction
\cite{Yue_Cao_2019,Dai2022}; however, posterior inference
of DM-spike parameters with eccentric EMRI waveforms remains less developed.
In this work, we therefore incorporate the leading dissipative effects of a
DM spike into an eccentric analytic-kludge (AK) waveform model and combine
FIM diagnostics with Markov-chain Monte Carlo (MCMC) sampling to investigate
the measurability of the DM-spike slope \(\alpha_{\rm DM}\), with particular
attention to its correlation with the orbital eccentricity.

The present analysis is based on the AK waveform model
\cite{BarackCutler2004}. Although more accurate EMRI waveform models are
available~\cite{Babak2007,Chua2017,DrascoHughes2006}, the AK
framework provides an efficient tool for exploring environmental effects and
performing computationally intensive Bayesian inference. The results should
therefore be regarded as a proof-of-principle study rather than a precision
forecast for future LISA observations. In this work, we construct eccentric AK waveforms for EMRIs embedded in a DM spike by including DF and accretion in the secular evolution of the orbital frequency and eccentricity. We first investigate the resulting modifications to the orbital evolution and waveform, and then perform MCMC parameter estimation to assess the measurability of the DM-spike parameters and their
correlations with the intrinsic source parameters. The remainder of this paper
is organized as follows. In Sec.~\ref{sec:EMRI}, we present the EMRI
orbital-evolution model in a DM environment and its implementation within the
AK framework. In Sec.~\ref{sec:Re and dis}, we compare waveforms with and
without DM, present parameter-recovery results for different plunge eccentricities, and analyze the correlations between the DM and intrinsic
source parameters. Finally, Sec.~\ref{sec:conclusion} summarizes our conclusions.

\section{Method}\label{sec:EMRI}
In this section, we briefly introduce the methodology of describing EMRIs immersed in a DM distribution, the analytic-kludge waveform model corrected by the DF effect of DM, and the parameter estimation based on MCMC.   

\subsection{DM spike density distribution and its construction}
\label{subsec:dm_spike_density}
We assume the DM distribution surrounding the central massive black
hole (MBH) as an adiabatically grown density spike. Following the standard
prescription for spike formation, the initial DM halo is
described by a power-law cusp~\cite{Navarro1997},
\begin{equation}
\rho_i(r)  = \rho_0 \left(\frac{r_0}{r}\right)^{\gamma},
\end{equation}
where \(\rho_0\) and \(r_0\) denote the characteristic density and scale
radius of the halo, respectively, and \(\gamma\) determines the inner
density slope. These quantities are fixed following the prescription of
Ref.~\cite{PhysRevD.99.043533}, which relates the properties of the host
halo to the mass and velocity dispersion of the central MBH system.
Assuming that the MBH grows adiabatically at the center of the halo, the
surrounding DM responds to the gradually deepening gravitational potential
and develops a steeper and more centrally concentrated density profile,
commonly referred to as a DM spike~\cite{Gondolo1999,Sadeghian2013,Ferrer2017}.

Incorporating relativistic phase-space corrections appropriate for a Schwarzschild background geometry, the DM spike density profile is phenomenologically modeled as follows ~\cite{Gondolo1999,Sadeghian2013,Eda:2013gg,Eda:2014kra,Hannuksela2020,Li2022,Montalvo2024}:
\begin{equation} 
\rho_{\rm DM}(r) = \rho_{\rm sp} \left(1-\frac{2R_s}{r}\right)^3 \left(\frac{r_{\rm sp}}{r}\right)^{\alpha_{\rm DM}}\;.
\end{equation}
Here, $\rho_{\rm sp}=\rho_0\left(\frac{r_0}{r_{\rm sp}}\right)^\gamma$ is the spike density normalization and $r_{\rm sp}=\beta_\gamma r_0
    \left(\frac{M}{\rho_0 r_0^3}\right)^{1/(3-\gamma)}$ parameterizes the characteristic spatial extent of the spike, which depends on the detailed black-hole formation history and subsequent dynamical evolution of the galactic nucleus ~\cite{Ullio2001,Merritt2002,Bertone2005}. An initial cusp with logarithmic slope $\gamma$ is transformed into a spike with slope $\alpha_{\rm DM}=(9-2\gamma)/(4-\gamma)$. For an initial range of $\gamma \in [0, 2]$, the spike slope lies tightly within $\alpha_{\rm DM} \in [2.25, 2.5]$. Here, $R_s=2GM/c^2$ is the Schwarzschild radius of the central MBH. The factor $(1-2R_s/r)^3$ enforces a sharp physical suppression of the DM density as $r \to 2R_s$, reflecting the absence of stable particle orbits strictly near the adopted inner cutoff ~\cite{Sadeghian2013,Ferrer2017,Xu2021}, and the profile is defined over the radial domain $r_{\min} \le r \le r_{\rm sp}$.

\subsection{DM-induced modifications to the secular orbital evolution}
\label{subsec:environmental_perturbations}

For an EMRI, the orbital motion occurs on a timescale \(T_{\rm orb}\) much shorter than the secular-evolution timescale \(T_{\rm sec}\). This separation of timescales motivates the AK description, in which the inspiral is represented as a sequence of instantaneous Keplerian ellipses whose orbital elements evolve slowly under radiation reaction and environmental perturbations~\cite{BarackCutler2004,Zi2023}. We retain the leading dissipative effects associated with the DM
environment, namely DF and accretion-induced drag.
These effects modify the secular evolution of the orbital frequency and
eccentricity and consequently generate an accumulated GW dephasing.
Conservative corrections arising from the gravitational potential of the
DM spike are neglected in the present analysis
~\cite{Eda:2014kra,Dai2022,Montalvo2024}.

At each instant, the perturbed trajectory is described by an osculating
Keplerian ellipse having the same position and velocity as the true orbit.
The radial motion is parameterized by the true anomaly \(\phi\),
measured from periapsis,
\begin{equation}
r(\phi)=\frac{p}{1+e\cos\phi},
\label{eq:rpsi_DM}
\end{equation}
where \(p\) is the semilatus rectum and \(e\) is the orbital eccentricity.
The orbital elements \((p,e,\omega)\), with \(\omega\) denoting the argument
of periapsis, are slowly varying functions of time.

Let \(\mathbf r=r\mathbf n\) and
\(\mathbf v=\dot{\mathbf r}\). The relative acceleration can be written as
\begin{equation}
\mathbf a=
-\frac{GM}{r^2}\mathbf n
+\mathbf f_{\rm other}
+\mathbf f_{\rm DM},
\label{eq:A1_like}
\end{equation}
where \(\mathbf f_{\rm other}\) contains GW radiation reaction and any
additional perturbing accelerations, while \(\mathbf f_{\rm DM}\) denotes
the dissipative acceleration produced by the DM environment. Introducing
the comoving orbital basis
\((\mathbf n,\mathbf k,\mathbf e_z)\), the perturbing acceleration is
decomposed as
\begin{equation}
\mathbf f=R\,\mathbf n+S\,\mathbf k+W\,\mathbf e_z ,
\label{eq:A2_like}
\end{equation}
where \(R\), \(S\), and \(W\) are the radial, tangential, and normal components, respectively. These components determine the evolution of the osculating orbital elements through the Gauss planetary equations~\cite{2014Gravity,DiCarlo2021}.
The dissipative acceleration produced by the DM is modeled as a
velocity-dependent drag force. Following the standard treatment of
DF and its application to compact-object inspirals in DM
spikes~\cite{Chandrasekhar1943,Eda:2014kra}, we write
\begin{equation}
\mathbf f_{\rm DM}
=
-\frac{4\pi G^2\mu\,\rho_{\rm DM}(r)}{v^3}
\left(I_v+I_\lambda\right)\mathbf v .
\label{eq:f_DM_drag}
\end{equation}
Here \(I_v\) characterizes the DF contribution, for which we adopt
\(I_v=3\) following Refs.~\cite{Eda:2014kra,Yue_Cao_2019}, while
\(I_\lambda=1\) describes the accretion-induced drag in the adopted
cold-medium Bondi--Hoyle--Lyttleton approximation
~\cite{Bondi1944,Edgar2004}. The orbital velocity is
\begin{equation}
\mathbf v=
\sqrt{\frac{GM}{p}}
\left[
e\sin\phi\,\mathbf n+
(1+e\cos\phi)\mathbf k
\right].
\label{eq:v_decomposition_osculating}
\end{equation}
The projections of the DM drag force are therefore
\begin{align}
R_{\rm DM}
&=
-\frac{4\pi G^2\mu\,\rho_{\rm DM}(r)}{v^3}
(I_v+I_\lambda)
\sqrt{\frac{GM}{p}}\,e\sin\phi ,
\\
S_{\rm DM}
&=
-\frac{4\pi G^2\mu\,\rho_{\rm DM}(r)}{v^3}
(I_v+I_\lambda)
\sqrt{\frac{GM}{p}}\,(1+e\cos\phi),
\\
W_{\rm DM}&=0 .
\label{eq:RSW_DM_drag}
\end{align}
The accumulated change of an orbital element \(K\) over one radial period
is obtained by integrating over the true anomaly. We define
\begin{equation}
\left\langle\frac{dK}{d\phi}\right\rangle_\phi
=
\frac{1}{2\pi}
\int_0^{2\pi}
\frac{dK}{d\phi}\,d\phi .
\label{eq:orbital_average}
\end{equation}
Since the orbital period is \(T_{\rm orb}=1/\nu\), the corresponding
secular time derivative is
\begin{equation}
\left\langle\dot K\right\rangle
=2\pi\nu\left\langle\frac{dK}{d\phi}\right\rangle_\phi .
\label{eq:phase_to_time_average}
\end{equation}
Using Kepler's law,
\begin{equation}
p=(GM)^{1/3}(1-e^2)(2\pi\nu)^{-2/3},
\end{equation}
or equivalently
\begin{equation}
\nu=
\frac{1}{2\pi}
\sqrt{\frac{GM}{p^3}}\,
(1-e^2)^{3/2},
\label{eq:nu_p_e}
\end{equation}
its derivation satisfies
\begin{equation}
\left\langle\frac{d\nu}{d\phi}\right\rangle_\phi
=\nu\left[
-\frac{3}{2p}
\left\langle\frac{dp}{d\phi}\right\rangle_\phi
-\frac{3e}{1-e^2}
\left\langle\frac{de}{d\phi}\right\rangle_\phi
\right]\;,
\label{eq:avg_dnu_dphi}
\end{equation}
we can obtain two evolution equations of orbital frequency and eccentricity modified by DM as following
\begin{subequations}\label{eq:secular_nu_e}
\begin{align}
\left\langle\dot{\nu}\right\rangle_{\rm DM}
&=
2\pi\nu^2\mathcal C_{\rm DM}
\left[
\frac{3}{2}\mathcal I_0
+\frac{3e}{1-e^2}\mathcal I_1
\right],
\label{eq:secular_dnu_dt}
\\[1ex]
\left\langle\dot e\right\rangle_{\rm DM}
&=
-(2\pi\nu)\mathcal C_{\rm DM}\mathcal I_1 \;,
\label{eq:secular_de_dt}
\end{align}
\end{subequations}
where 
\begingroup
\begin{equation}
\begin{aligned}
\mathcal C_{\rm DM}
&=
\frac{4\mu\rho_{\rm sp}r_{\rm sp}^{\alpha_{\rm DM}}}{M^2}
(I_v+I_\lambda)
(GM)^{(3-\alpha_{\rm DM})/3}
\\
&\quad\times
(1-e^2)^{3-\alpha_{\rm DM}}
(2\pi\nu)^{-2(3-\alpha_{\rm DM})/3}\;,
\end{aligned}
\label{eq:C_DM}
\end{equation}
\endgroup
with two quantities
\begin{align}
\mathcal I_0
&=
\frac{1}{GM(1-e^2)^3}
\int_0^{2\pi}
\frac{\mathcal Q(\phi)^3\,d\phi}
{\mathcal P(\phi)^{3/2}\mathcal R(\phi)^{2-\alpha_{\rm DM}}},
\label{eq:IDMzero}
\\
\mathcal I_1
&=\frac{1}{GM(1-e^2)^3}
\int_0^{2\pi}
\frac{(e+\cos\phi)\mathcal Q(\phi)^3\,d\phi}
{\mathcal P(\phi)^{3/2}\mathcal R(\phi)^{2-\alpha_{\rm DM}}}\;,\label{eq:IDMone}
\end{align}
with
\begingroup
\begin{equation}
\begin{gathered}
\mathcal Q(\phi)=
(GM)^{1/3}(1-e^2)
-2R_s(2\pi\nu)^{2/3}(1+e\cos\phi),\\
\mathcal P(\phi)=1+2e\cos\phi+e^2,\qquad
\mathcal R(\phi)=1+e\cos\phi.
\end{gathered}
\label{eq:QPR_definitions}
\end{equation}
\endgroup
The term proportional to \(\mathcal I_1\) in
Eq.~\eqref{eq:secular_dnu_dt} originates from the eccentricity dependence
of the Keplerian relation \(p(\nu,e)\). It must therefore be retained
consistently when converting the DM-induced evolution of \(p\) and \(e\)
into the corresponding evolution of the orbital frequency.

\subsection{AK waveform model modified by DM spikes}
The AK framework~\cite{BarackCutler2004} is
particularly useful in this context because it captures the main
phenomenological features of EMRI signals, including the secular evolution of
the orbital frequency and eccentricity, while remaining computationally efficient. Although
more accurate waveform models based on gravitational self-force calculations
and fully relativistic treatments are being actively developed
\cite{Babak2007,Chua2017,DrascoHughes2006}, the AK model remains well
suited to exploratory studies of additional physical effects and to
large-scale parameter-space investigations.

In this section, we construct AK waveforms for an EMRI evolving inside a DM
minispike. Following the standard AK prescription, the instantaneous orbital
motion is described by an osculating Keplerian ellipse, while the orbital
elements and precession angles evolve adiabatically under the combined action
of conservative relativistic precession and dissipative radiation reaction.
Relative to the vacuum AK model, we modify only the dissipative evolution
equations for the orbital frequency \(\nu\) and eccentricity \(e\) by adding
the DM-induced contributions derived above. Accordingly, the effect of the DM environment enters the waveform primarily
through its modification of the secular orbital evolution
\begin{equation}
\frac{dX}{dt}
=
\left(\frac{dX}{dt}\right)_{\rm DM}
+
\left(\frac{dX}{dt}\right)_{\rm other},
\qquad
X\in\{\phi,\nu, e,\alpha,\tilde{\gamma}\}.
\label{eq:sum_DM_GW}
\end{equation}
We combine these leading order
corrected equations with those higher-order PN equations
in the original AK model ~\cite{BarackCutler2004}. We define the PN expansion parameter as $x=\left(\frac{2\pi GM\nu}{c^3}\right)^{1/3}$. The complete AK evolution equations used in this work are
\begin{align}
\dot{\Phi}
&=2\pi\nu ,
\label{eq:dphi_dt}
\\
\dot{\nu}
&=
\frac{96}{10\pi}
\frac{c^6\mu}{G^2M^3}
\PNx^{11}(1-e^2)^{-9/2}
\notag\\
&\quad\times
\Bigg[
(1-e^2)
\left(1+\frac{73}{24}e^2+\frac{37}{96}e^4\right)
\notag\\
&\qquad
+\PNx^2
\left(\frac{1273}{336}-\frac{2561}{224}e^2
-\frac{3885}{128}e^4-\frac{13147}{5376}e^6\right)
\notag\\
&\qquad
-\PNx^3
\frac{cS}{GM^2}
\cos\lambda\,
(1-e^2)^{-1/2}
\notag\\
&\qquad\times
\left(
\frac{73}{12}
+\frac{1211}{24}e^2
+\frac{3143}{96}e^4
+\frac{65}{64}e^6
\right)
\Bigg]
\notag\\
&\quad
+2\pi\nu^2
\frac{4\mu\rhosp\rsp^{\alpha_{\rm DM}}}{M^2}
\left(I_v+I_{\lambda}\right)
\notag\\
&\quad\times
(GM)^{(3-\alpha_{\rm DM})/3}
(1-e^2)^{3-\alpha_{\rm DM}}
\notag\\
&\quad\times
(2\pi\nu)^{-2(3-\alpha_{\rm DM})/3}
\left[\frac{3}{2}\IDMzero
+\frac{3e}{1-e^2}\IDMone\right],
\label{eq:df_dtAK}
\\
\dot{e}
&=
-\frac{e}{15}
\frac{c^3\mu}{GM^2}
\PNx^8(1-e^2)^{-7/2}
\notag\\
&\quad\times
\Bigg[
(1-e^2)(304+121e^2)(1+12\PNx^2)
\notag\\
&\qquad
-\frac{\PNx^2}{56}
\left(133640+108984e^2-25211e^4\right)
\Bigg]
\notag\\
&\quad
+e\frac{c^3\mu}{GM^2}
\frac{cS}{GM^2}
\cos\lambda\,
\PNx^{11}(1-e^2)^{-4}
\notag\\
&\quad\times
\left(\frac{1364}{5}+\frac{5032}{15}e^2
+\frac{263}{10}e^4\right)
\notag\\
&\quad
-(2\pi\nu)
\frac{4\mu\rhosp\rsp^{\alpha_{\rm DM}}}{M^2}
\left(I_v+I_{\lambda}\right)
\notag\\
&\quad\times
(GM)^{(3-\alpha_{\rm DM})/3}
(1-e^2)^{3-\alpha_{\rm DM}}
\notag\\
&\quad\times
(2\pi\nu)^{-2(3-\alpha_{\rm DM})/3}
\IDMone ,
\label{eq:e_t}
\\
\dot{\tilde{\gamma}}
&=
6\pi\nu
\PNx^2
(1-e^2)^{-1}
\left[
1+\frac{\PNx^2}{4(1-e^2)}
(26-15e^2)
\right]
\notag\\
&\quad
-12\pi\nu
\frac{cS}{GM^2}
\cos\lambda\,\PNx^3
(1-e^2)^{-3/2},
\label{eq:gamma_t}
\\
\dot{\alpha}
&=4\pi\nu
\frac{cS}{GM^2}
\PNx^3(1-e^2)^{-3/2}.
\label{eq:alpha_t}
\end{align}
Following the conventions in the Ref.~\cite{BarackCutler2004}, we extend the equations for $\dot{\nu}$ and $\dot{e}$ to 3.5 PN order, while the equations for $\dot{\tilde{\gamma}}$ ($\tilde{\gamma}$ is the angle between $\hat{L}\times\hat{S}$ and the periapsis, where $\hat{L}$ is the unit vector of the orbital angular momentum and $\hat S$ is the unit vector along the spin of the central massive black hole.) and $\dot{\alpha}$ are extended to 2 PN order which describes the Lense--Thirring precession of the orbital plane.  $\Phi$ is the mean phase, $\lambda$ denotes the inclination angle of the orbital plane relative to the spin direction of the EMRI, and $S/M^2$ represents the dimensionless spin parameter.

In the weak-field and slow-motion regime, the transverse--traceless (TT)
metric perturbation measured at a luminosity distance \(D\) is given by the
quadrupole formula
\begin{equation}
h_{ij}^{\rm TT}
=
\frac{2G}{c^4D}
\left(
P_{ik}P_{jl}
-\frac{1}{2}P_{ij}P_{kl}
\right)
\ddot{I}^{kl},
\label{eq:quadrupole}
\end{equation}
where
\begin{equation}
P_{ij}=\delta_{ij}-\hat{n}_i\hat{n}_j
\end{equation}
is the projection tensor transverse to the line of sight
\(\hat{\boldsymbol n}\). For the DM-dressed EMRI considered here, the
mass quadrupole is determined by the orbital motion of the secondary,
\begin{equation}
I_{ij}(t)
=
\mu(t)\,r^2(t)\,
\hat{r}_i(t)\hat{r}_j(t),
\label{eq:quadrupole_orbit}
\end{equation}
where \(\hat{\boldsymbol r}\) denotes the instantaneous radial unit vector
in the orbital plane and \(\mu(t)\) is the mass of the inspiralling
secondary object. The accretion of DM halo leads to a slowly varying secondary
mass, the changing rate is modeled as
\begin{equation}
\dot{\mu}(t)
=
4\pi G^2 I_{\lambda}
\frac{\mu^2(t)\rho_{\rm DM}}
{\left[v^2(t)+c_s^2\right]^{3/2}},
\label{eq:DM_accretion_mass}
\end{equation}
following the Bondi--Hoyle--Lyttleton prescription
\cite{Bondi1944,Edgar2004}. For the parameter ranges considered in this
work, however, the accumulated fractional mass change
\(\Delta\mu/\mu\) remains very small. We therefore neglect its direct
contribution to the waveform amplitude, while retaining the effect of
accretion on the secular orbital evolution as described above.

For an eccentric Keplerian orbit, the quadrupole moment can be decomposed
into harmonics of the orbital motion,
\begin{equation}
I_{ij}(t)=\sum_{n=1}^{\infty} I_{ij,n}(t).
\end{equation}
It is convenient to introduce the combinations
\begin{equation}
\begin{aligned}
a_n&=\frac{1}{2}
\left(\ddot I_{11,n}-\ddot I_{22,n}\right),
&
b_n&=\ddot I_{12,n},
\\
c_n&=\frac{1}{2}
\left(\ddot I_{11,n}+\ddot I_{22,n}\right),
\end{aligned}
\label{eq:abc_harmonics}
\end{equation}
whose explicit expressions in terms of Bessel functions are those of the
standard AK waveform model~\cite{BarackCutler2004}.

We define the inclination angle \(\iota\) through
\begin{equation}
\cos\iota=\hat{\boldsymbol L}\cdot\hat{\boldsymbol n},
\end{equation}
where \(\hat{\boldsymbol L}\) is the unit orbital-angular-momentum vector.
The angle \(\gamma(t)\) specifies the orientation of the periapsis within
the orbital plane relative to the projection of
\(\hat{\boldsymbol n}\) onto that plane. The contribution of the \(n\)th
orbital harmonic to the two GW polarizations is then
\begin{subequations}\label{eq:hpol_sum}
\begin{align}
h^{(+)}_{n}
=&
-\frac{G}{c^{4}D}
\Big\{
(1+\cos^2\iota)
\left[
a_n\cos(2\gamma)-b_n\sin(2\gamma)
\right]
\\ \nonumber &+c_n\sin^2\iota
\Big\},
\\
h^{(\times)}_{n}
&=
\frac{2G}{c^{4}D}
\cos\iota
\left[
b_n\cos(2\gamma)+a_n\sin(2\gamma)
\right].
\end{align}
\end{subequations}
These expressions explicitly separate the rapidly varying orbital
harmonics, encoded in \(a_n\), \(b_n\), and \(c_n\), from the more slowly
varying geometrical modulation through \(\iota\) and \(\gamma\).
The total time-domain polarizations are obtained by summing over the
orbital harmonics,
\begin{equation}
h_{+}(t)=\sum_{n=1}^{\infty}h^{(+)}_n(t),
\qquad
h_{\times}(t)=\sum_{n=1}^{\infty}h^{(\times)}_n(t).
\label{eq:total_polarizations}
\end{equation}

To perform parameter estimation, the source-frame polarizations need to be
projected onto the response of a space-based GW detector. Although the
general formalism applies to detectors such as LISA
\cite{Amaro_Seoane_2017}, Taiji~\cite{Taiji_Collaboration_2021}, and
TianQin~\cite{Luo_2016}, in the present work we mainly focus to LISA.
Within the low-frequency approximation, the responses of the two
independent Michelson-like channels are written as
\begin{equation}
h_{I,II}(t)
=
\frac{\sqrt{3}}{2}
\left[
F^{+}_{I,II}(t)h_{+}(t)
+
F^{\times}_{I,II}(t)h_{\times}(t)
\right],
\label{eq:detector_response}
\end{equation}
where \(F^{+,\times}_{I,II}(t)\) are the time-dependent antenna-pattern
functions~\cite{Cutler1998LISA}. Throughout this work, we adopt the
low-frequency LISA response used in the standard AK framework
\cite{BarackCutler2004}.

\subsection{Parameter-estimation methods}
\label{subsec:parameter_estimation}

We constrain the source and DM parameters within a Bayesian
framework. The data streams in the LISA-like detectors are modeled as
\begin{equation}
s_{\alpha}(t)  =  h_{\alpha}(t;\boldsymbol{\theta}) +  n_{\alpha}(t),
\label{eq:data_model}
\end{equation}
where \(h_{\alpha}(t;\boldsymbol{\theta})\) denotes the GW signal in the
detector channel \(\alpha\), specified by the parameter vector
\(\boldsymbol{\theta}\), and \(n_{\alpha}(t)\) is the detector noise.
Bayes's theorem gives
\begin{equation}
    p(\boldsymbol{\theta}|s)
    =
    \frac{
    p(s|\boldsymbol{\theta})\,p(\boldsymbol{\theta})
    }{
    p(s)
    },
\label{eq:bayes_theorem}
\end{equation}
where \(p(\boldsymbol{\theta}|s)\) is the posterior probability density,
\(p(s|\boldsymbol{\theta})\) is the likelihood,
\(p(\boldsymbol{\theta})\) denotes the prior, and \(p(s)\) is the Bayesian
evidence. Since the evidence is independent of
\(\boldsymbol{\theta}\), parameter inference is based on
\begin{equation}
    p(\boldsymbol{\theta}|s)
    \propto
    p(s|\boldsymbol{\theta})\,p(\boldsymbol{\theta}).
\label{eq:posterior_proportional}
\end{equation}

Assuming stationary Gaussian noise, we define the noise-weighted inner
product between two signals \(a\) and \(b\) as
\begin{equation}
    (a|b)
    =
    4\,{\rm Re}
    \sum_{\alpha}
    \int_{0}^{\infty}
    \frac{
    \tilde a_{\alpha}^{*}(f)\tilde b_{\alpha}(f)
    }{
    S_{n,\alpha}(f)
    }\,df ,
\label{eq:inner_product}
\end{equation}
where \(S_{n,\alpha}(f)\) is the one-sided noise power spectral density
of channel \(\alpha\)
\cite{Cutler_1994,finn2000gravitational}. For the two independent LISA
channels adopted in this work, the sum runs over
\(\alpha=I,II\). The Gaussian likelihood is then
\begin{equation}
    p(s|\boldsymbol{\theta})
    \propto
    \exp\left[
    -\frac{1}{2}
    \left(
    s-h(\boldsymbol{\theta})
    \middle|
    s-h(\boldsymbol{\theta})
    \right)
    \right].
\label{eq:gaussian_likelihood}
\end{equation}

The complete AK waveform depends on a relatively large set of intrinsic,
extrinsic, and phase parameters~\cite{BarackCutler2004}.

For the MCMC analyses, the sampler evolves the full 14-dimensional parameter
vector listed in Table~\ref{tab:mcmc_priors_full} and given explicitly in
Eq.~\eqref{sources:emri}. 
Since the main purpose of the present analysis is to investigate the measurability of the
DM-spike profile and its degeneracy with the intrinsic orbital dynamics, we
mainly show the sampled parameter space to
    \begin{equation}
    \boldsymbol{\theta}
    =
    \left\{
    \ln M,\,
    \ln\eta,\,
    \kai,\,
    e_{\rm LSO},\,
    \alpha_{\rm DM}
    \right\},
    \label{eq:sampled_parameter_space}
    \end{equation}
where \(\eta=\mu/M\) denotes the mass ratio and
\(\alpha_{\rm DM}\) characterizes the slope of the DM spike.
Logarithmic variables are adopted for the mass parameters to improve the
numerical conditioning of the sampling and to naturally characterize
fractional variations in the EMRI masses. Unless otherwise specified,
uniform priors are imposed on the sampled parameters.

The posterior in Eq.~\eqref{eq:posterior_proportional} is explored with the affine-invariant ensemble sampler implemented in the \texttt{emcee} package~\cite{emcee}. For an ensemble of walkers, the stretch-move proposal updates a walker \(\boldsymbol{\theta}_k\) relative to another walker \(\boldsymbol{\theta}_j\) according to
\begin{equation}
\boldsymbol{\theta}'_k
=
\boldsymbol{\theta}_j
+z\left(\boldsymbol{\theta}_k-\boldsymbol{\theta}_j\right),
\label{eq:emcee_stretch_move}
\end{equation}
where \(z\) is drawn from the standard stretch-move distribution. The proposed move is accepted with probability
\begin{equation}
A(\boldsymbol{\theta}'_k|\boldsymbol{\theta}_k)
=
\min\left[
1,
z^{N_{\rm par}-1}
\frac{
p(s|\boldsymbol{\theta}'_k)p(\boldsymbol{\theta}'_k)
}{
p(s|\boldsymbol{\theta}_k)p(\boldsymbol{\theta}_k)
}
\right].
\label{eq:mh_acceptance}
\end{equation}
Here \(N_{\rm par}=14\) is the dimension of the sampled parameter space. After discarding the burn-in portion and verifying convergence, the retained
samples are used to construct marginalized posterior distributions and to
quantify parameter uncertainties and degeneracies.

For comparison with the full Bayesian analysis, we also employ the FIM as a local approximation to the likelihood in the
high-signal-to-noise-ratio regime. Expanding the waveform about a fiducial
parameter point \(\boldsymbol{\theta}_0\),
\begin{equation}
    h(\boldsymbol{\theta})
    \simeq
    h(\boldsymbol{\theta}_0)
    +
    \Delta\theta_i\,\partial_i h,
    \qquad
    \Delta\theta_i
    =
    \theta_i-\theta_{0,i},
\label{eq:waveform_linear_expansion}
\end{equation}
where
\(\partial_i h\equiv\partial h/\partial\theta_i\), the FIM is
defined by
\begin{equation}
    \Gamma_{ij}
    =
    \left(
    \partial_i h
    \middle|
    \partial_j h
    \right)_{\boldsymbol{\theta}=\boldsymbol{\theta}_0}.
\label{eq:fisher_matrix}
\end{equation}
For forecasting purposes, and upon averaging over Gaussian noise
realizations, the likelihood near \(\boldsymbol{\theta}_0\) takes the
quadratic form
\begin{equation}
    p(s|\boldsymbol{\theta})
    \propto
    \exp\left[
    -\frac{1}{2}
    \Gamma_{ij}
    \Delta\theta_i\Delta\theta_j
    \right].
\label{eq:local_gaussian_likelihood}
\end{equation}
Provided that the signal-to-noise ratio is sufficiently high and the priors
are broad compared with the likelihood, the covariance matrix can be
approximated as
\begin{equation}
    \Sigma_{ij}
    \simeq
    \left(\Gamma^{-1}\right)_{ij}.
\label{eq:fisher_covariance}
\end{equation}
The corresponding \(1\sigma\) statistical uncertainties and linear
correlation coefficients are
\begin{equation}
\sigma_{\theta_i}   =   \sqrt{\Sigma_{ii}},    \qquad
r(\theta_i,\theta_j)
=\frac{\Sigma_{ij}}{\sigma_{\theta_i}\sigma_{\theta_j}}\;,
\label{eq:fisher_errors_correlations}
\end{equation}
where the coefficient between $\theta_i$ and $\theta_j$ reflects the dependent correlation in the FIM analysis.

In the following, the FIM is used primarily as a computationally efficient
diagnostic of local parameter uncertainties and linear correlations, whereas
the MCMC analysis provides the main characterization of the posterior,
including possible asymmetries, non-Gaussian structures, and nonlinear
degeneracies between the DM-spike and intrinsic EMRI parameters.

\section{Results and discussion}\label{sec:Re and dis}
In this section, we present the numerical results of our analysis. We begin by
comparing EMRI waveforms generated in vacuum and in the presence of a
DM environment, thereby quantifying the impact of the latter on the
GW signal. We then assess the constraints that LISA could place
on the source and DM parameters for a set of representative
detectable EMRI systems.
\begin{figure*}
\centering    
\includegraphics[width=0.97\textwidth]{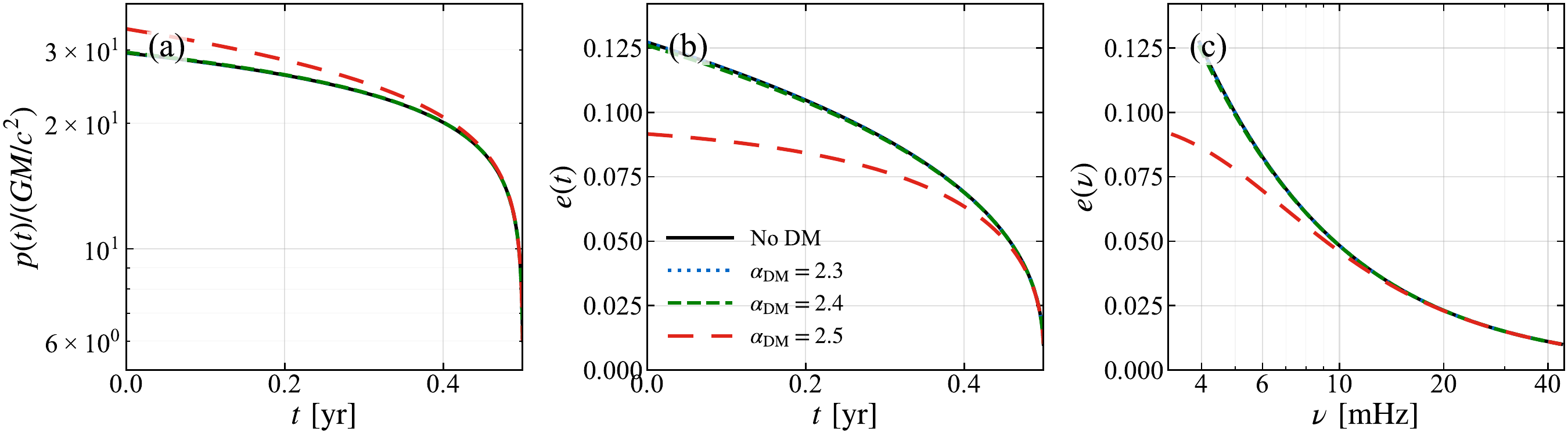}
\caption{Orbital evolution for different DM-spike
slopes \(\alpha_{\rm DM}\in\{2.3,2.4,2.5\}\), fixing the eccentricity at the LSO to \(e_{\rm LSO}=0.01\). Panel (a) shows the semilatus-rectum evolution, \(p(t)/(GM/c^2)\), panel (b) shows the eccentricity evolution \(e(t)\), and panel (c) shows the eccentricity as a function of orbital frequency, \(e(\nu)\). Since all trajectories are evolved to the same terminal LSO condition, they coincide at the endpoint, while their earlier differences reflect the accumulated effect of the DM environment on the orbital
evolution. The remaining source parameters are
\(M=5.0\times10^{4}M_{\odot}\), \(\mu=10M_{\odot}\),
\(\chi=0.1\), and \(t_c=0.5~{\rm yr}\).
The terminal orbital frequency is determined by
\(\nu_{\rm LSO}=c^{3}(2\pi GM)^{-1}
[(1-e_{\rm LSO}^{2})/(6+2e_{\rm LSO})]^{3/2}\).}\label{Fig:evolution}
\end{figure*}

\subsection{Orbital and waveform imprints of the DM spike}
\begin{figure*}[t]
\centering    
\includegraphics[width=1.0\textwidth]{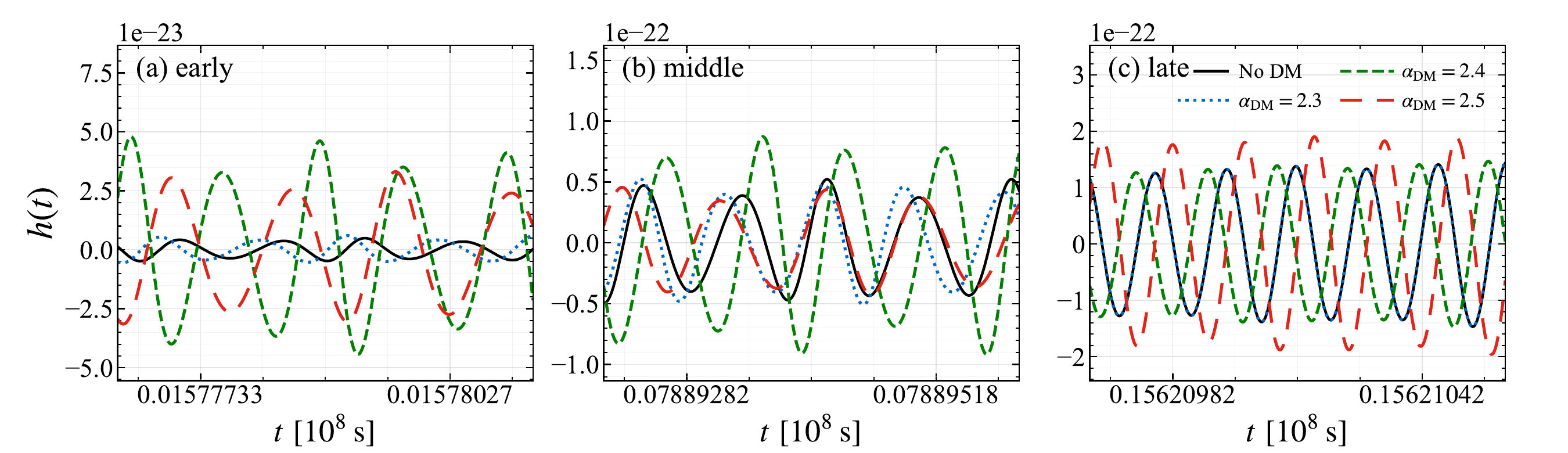}
\caption{The time-domain GW waveforms \(h(t)\) are shown for
three representative intervals: \(t/(10^8\,{\rm s})\in[0.01577635, 0.01578125]\)
(early, left panel), \([0.07889204, 0.07889596]\) (middle panel), and
\([0.15620962, 0.15621062]\) (late, right panel). The curves compare the vacuum waveform with the DM-spike cases \(\alpha_{\rm DM}=2.3\), \(2.4\), and \(2.5\).} \label{fig:ht}
\end{figure*}

To illustrate the impact of the DM environment on the EMRI orbital
evolution, Fig.~\ref{Fig:evolution} shows the time evolution of the orbital
semilatus rectum and eccentricity, together with the eccentricity-frequency
relation, for three representative DM-spike
slopes,
\(\alpha_{\rm DM}\in\{2.3,2.4,2.5\}\), together with the corresponding
vacuum evolution. To isolate the effect of the DM distribution, we vary
only \(\alpha_{\rm DM}\) while keeping all other source parameters fixed.
The black solid curve denotes the vacuum case, whereas the blue dotted,
green dashed, and red dashed curves correspond to
\(\alpha_{\rm DM}=2.3\), \(2.4\), and \(2.5\), respectively. The
remaining parameters are fixed to 
$\{t_c=0.5~{\rm yr},
z=0.1,
\mu=10\,M_{\odot},
M=5.0\times10^{4}\,M_{\odot},
e_{\rm LSO}=0.01, \nu_{\rm LSO}=43.73~{\rm mHz}, \lambda=40^\circ\}$
together with
$\{
\tilde{\gamma}_{\rm LSO}
=\alpha_{\rm LSO}
=\theta_S
=\phi_S
=\phi_K
=\Phi_{\rm LSO}
=\frac{\pi}{3},
\theta_K=\frac{2\pi}{3}
\}$.
For a direct comparison, we impose the same plunging orbital conditions,
\(\nu_{\rm LSO}=43.73~{\rm mHz}\) and \(e_{\rm LSO}=0.01\), at the LSO and
evolve each system over the preceding 0.5 year inspiral. In all cases, the
eccentricity decreases monotonically toward the prescribed value
\(e_{\rm LSO}=0.01\) at plunge. However, the presence of DM can modify the rate
of eccentricity evolution and even increase the eccentricity~\cite{Yue_Cao_2019,Li2022}. For the same final
conditions and observation time, the inferred initial eccentricities
satisfy
\begin{equation}
e_{\rm ini}^{\alpha_{\rm DM}=2.5}
<
e_{\rm ini}^{\alpha_{\rm DM}=2.4}
<
e_{\rm ini}^{\alpha_{\rm DM}=2.3}
\lesssim
e_{\rm ini}^{\rm vac},\nonumber
\end{equation}
where the superscript ``$\alpha_{\rm DM}\in\{2.3,2.4,2.5\}$'' and ``$\rm vac$'' denote the DM-spike and vacuum cases, respectively, and the subscript ``ini'' denotes the value at the beginning of the evolution. Thus, a steeper DM spike produces a stronger modification on the secular evolution of orbital eccentricity. This behavior can be understood from the additional dissipative contribution of DF, which alters the relative rates of orbital-energy and angular-momentum loss, and thereby modifies the circularization of the eccentric orbit~\cite{Yue_Cao_2019,Li2022}. For the parameter range considered here,
the DM contribution reduces the rate of circularization relative to the
vacuum evolution; consequently, when all configurations are required to
reach the same \(e_{\rm LSO}\) after 0.5 years, the DM-dressed inspirals
start from progressively smaller eccentricities as
\(\alpha_{\rm DM}\) increases.

By solving the orbital-evolution equations in
Eqs.~\eqref{eq:dphi_dt}--Eq.~\eqref{eq:alpha_t} and substituting the resulting trajectories
into the waveform expressions in Eqs.~\eqref{eq:hpol_sum}, we obtain the
time-domain GW signals shown in Fig.~\ref{fig:ht}. For larger values of the
DM-spike slope, in particular for the case of \(\alpha_{\rm DM}=2.4\) and \(2.5\), the waveforms evolves the increasing phase differences relative to the corresponding
vacuum signal. The dephasing increases as \(\alpha_{\rm DM}\) is larger, reflecting
the stronger environmental dissipation in a denser DM spike. DF and accretion modify the secular evolution of the orbital frequency
and eccentricity, causing the DM-dressed inspiral to accumulate orbital phase
at a different rate from that in vacuum. As a result, the waveform phase departs from the vacuum prediction over the observation time.

The DM environment also modifies the waveform amplitude through its influence
on the instantaneous orbital frequency and eccentricity. Since the harmonic
amplitudes in the AK waveform depend on both quantities, changes in the
orbital evolution induced by DM lead to corresponding amplitude modulations.
For the configurations considered here, larger values of
\(\alpha_{\rm DM}\) produce more pronounced differences in both the phase
and amplitude relative to the vacuum waveform.


\begin{table*}[!t]
\centering
\caption{
Injected values and prior ranges of the parameters used in the
14-dimensional MCMC analysis. The parameters \(\phi_0\) and \(t_c\) are fixed.
}
\label{tab:mcmc_priors_full}
\scriptsize
\renewcommand{\arraystretch}{1.18}
\begin{ruledtabular}
\begin{tabular}{cccc}
Parameter & Injected value & Prior range & Meaning \\
\hline
\(\ln M\) & \(\ln(10^6M_\odot)\) &
\([\ln(0.5M_{\rm true}),\,\ln(2M_{\rm true})]\) &
central MBH mass \\

\(\ln\eta\) & \(\ln\eta_{\rm true}\) &
\([\ln(\eta_{\rm true}/100),\ln0.25]\)&
 mass ratio \\

\(\chi\) & \(0.1\) & \([0,1]\) &
central MBH spin \\

\(e_{\rm LSO}\) & \(0.6\) & \([0.5,0.7]\) &
eccentricity at the LSO \\

\(\alpha_{\rm DM}\) & \(2.4\) & \([2.25,2.5]\) &
DM spike slope \\

\(z\) & \(0.1\) & \([10^{-4},1]\) &
redshift \\

\(\lambda\) & \(\pi/6\) & \([0,\pi]\) &
spin--orbit inclination \\

\(\Phi_{\rm LSO}\) & \(0\) & \([-\pi,\pi]\) &
mean orbital phase at the LSO \\

\(\gamma_{\rm LSO}\) & \(0\) & \([-\pi,\pi]\) &
argument of periapsis at the LSO \\

\(\alpha_{\rm LSO}\) & \(0\) & \([-\pi,\pi]\) &
azimuth of orbital angular momentum \\

\(\theta_S\) & \(\pi/4\) & \([0,\pi]\) &
source polar angle \\

\(\phi_S\) & \(0\) & \([-\pi,\pi]\) &
source azimuthal angle \\

\(\theta_K\) & \(\pi/8\) & \([0,\pi]\) &
MBH-spin polar angle \\

\(\phi_K\) & \(0\) & \([-\pi,\pi]\) &
MBH-spin azimuthal angle \\

\(\phi_0\) & \(0\) &--- &
initial phase \\

\(t_c\) & \(3.127\times10^7\) &--- &inspiral time\\
\end{tabular}
\end{ruledtabular}
\end{table*}

\subsection{Posterior constraints and parameter correlations}
To obtain the robust constraints on the DM parameters and to assess the
consistency of two parameter-estimation methods, we compare the posterior
constraints obtained from the FIM with those from
a full Bayesian MCMC analysis for a representative
EMRI embedded in a DM spike. The Bayesian sampling is performed in Python
using the affine-invariant ensemble sampler implemented in the
\texttt{emcee} package~\cite{emcee}. We sample the 14-dimensional parameter
vector listed in Table~\ref{tab:mcmc_priors_full},
\begin{equation}
\begin{aligned}
\boldsymbol{\theta}
=\{
&\ln M,\,
\ln\eta,\,
\chi,\,
e_{\rm LSO},\,
\alpha_{\rm DM},\,
z,\,
\lambda,\,
\Phi_{\rm LSO},\,
\gamma_{\rm LSO},\,
\alpha_{\rm LSO},\,
\\ &\theta_S,\,
\phi_S,\,
\theta_K,\,
\phi_K
\},
\end{aligned}
\label{sources:emri}
\end{equation}
where \(\eta=\mu/M\) denotes the mass-ratio and
\(\chi=S/M^2\) is the dimensionless spin of the central black hole. The
initial orbital phase \(\phi_0\) and the coalescence time \(t_c\) are held
fixed. The injected parameters are
\begin{align}
M &= 10^6 M_\odot, &
\mu &= 10 M_\odot, &
\chi &= 0.1, &
e_{\rm LSO} &= 0.6, \nonumber\\
\alpha_{\rm DM} &= 2.4, &
z &= 0.1, &
\lambda &= \pi/6, &
t_c &= 1~ {\rm yr},\nonumber
\end{align}
with
\begin{equation}
\Phi_{\rm LSO}
=\gamma_{\rm LSO}
=\alpha_{\rm LSO}
=\phi_S
=\phi_K=0,
~
\theta_S=\pi/4,~\theta_K=\frac{\pi}{8}.
\nonumber
\end{equation}

The likelihood is evaluated in the frequency domain using the
noise-weighted inner product defined above. For each proposed parameter set,
the time-domain AK waveform is Fourier transformed into the frequency domain before evaluating the likelihood. We retain orbital
harmonics up to \(n_{\max}=21\) and adopt a sampling interval
\(dt=1\rm s\). When noise realizations are considered, complex Gaussian noise
is generated directly in the frequency domain with a variance determined by
the adopted one-sided LISA noise power spectral density.
For each analysis, we run four independent ensemble chains, each containing
48 walkers, for the 10000 MCMC steps, using four parallel worker processes. The
first 5000 steps of each chain are discarded as burn-in. Convergence is
assessed by comparing the independent ensembles using the Gelman--Rubin
diagnostic \(\hat{R}\)~\cite{GelmanRubin1992,Vehtari:2021}, together with an
inspection of the chain mixing before constructing the marginalized posterior
distributions. We required \(\hat{R}<1.01\) for every sampled parameter. If this condition was not satisfied, the   corresponding MCMC run was continued with additional steps until the convergence criterion was met~\cite{Vehtari:2021,universe11080259}. The retained post-burn-in samples are then used to determine credible intervals and parameter correlations. For the FIM, we determine finite-difference convergence by scanning different finite-difference step sizes and using higher-order finite-difference stencils. The numerical calculations are performed on a computing server equipped with an NVIDIA GeForce RTX 4090 GPU.

\begin{figure*}[!t]
    \centering
\includegraphics[width=\linewidth,trim=1.0 1.0 1.0 1.0,clip]{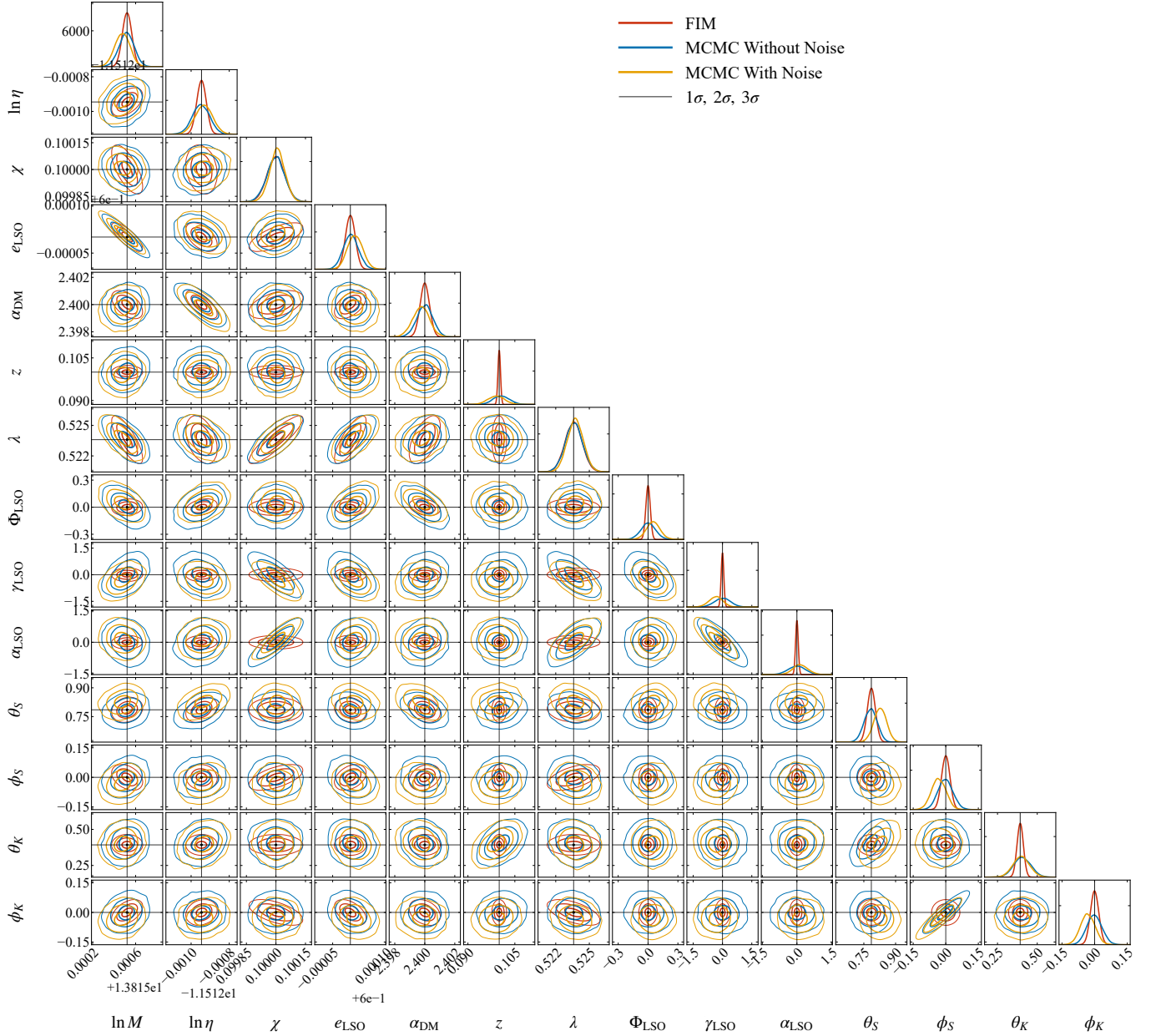}
\caption{
Comparison between the FIM computations and the MCMC posteriors
for a representative DM-dressed EMRI system. The injected source
parameters are \(M=10^{6}M_{\odot}\), \(\mu=10M_{\odot}\),
\(\chi=0.1\), \(e_{\rm LSO}=0.6\), \(\alpha_{\rm DM}=2.4\),
\(z=0.1\), and \(t_c=3.127\times10^{7}\,{\rm s}\). The other injected true values of parameters are listed in table~\ref{tab:mcmc_priors_full},
while \(\phi_0\) and \(t_c\) are fixed. The red contours show the local Gaussian approximation obtained from the FIM, whereas the blue and yellow contours show the MCMC posteriors for the zero-noise injection and for a realization including LISA detector noise, respectively. The three nested contours correspond to the \(1\sigma\), \(2\sigma\) and \(3\sigma\) levels shown in the corner plot.}\label{fig:fim-mc-noise}
\end{figure*}

Figure~\ref{fig:fim-mc-noise} presents the corner plot for the
14-dimensional parameter-estimation analysis. The red contours show the
local Gaussian approximation predicted by the FIM, while the blue and yellow contours represent the marginalized MCMC
posteriors obtained from the zero-noise injection and from a realization
including LISA detector noise, respectively. The contours correspond to the \(1\sigma\), \(2\sigma\), and \(3\sigma\) levels indicated in the figure. Uniform priors are adopted for all sampled parameters.
As shown in Fig.~\ref{fig:fim-mc-noise}, the FIM provides a reasonable
estimate of the local width and correlation structure of the posterior in
the vicinity of the injected parameters, particularly the intrinsic source parameters can be better-constrained. However, the full MCMC posterior also displays non-Gaussian features and parameter correlations, which illustrates the necessity of performing an MCMC analysis to characterize the full posterior structure. These results demonstrate that the FIM is useful as a fast local diagnostic, whereas a full Bayesian analysis is required to characterize the global posterior structure and to obtain more reliable constraints on the DM and EMRI parameters.

\begin{figure}[t]
\centering
\includegraphics[width=\linewidth,trim=1.0 1.0 1.0 1.0,clip]{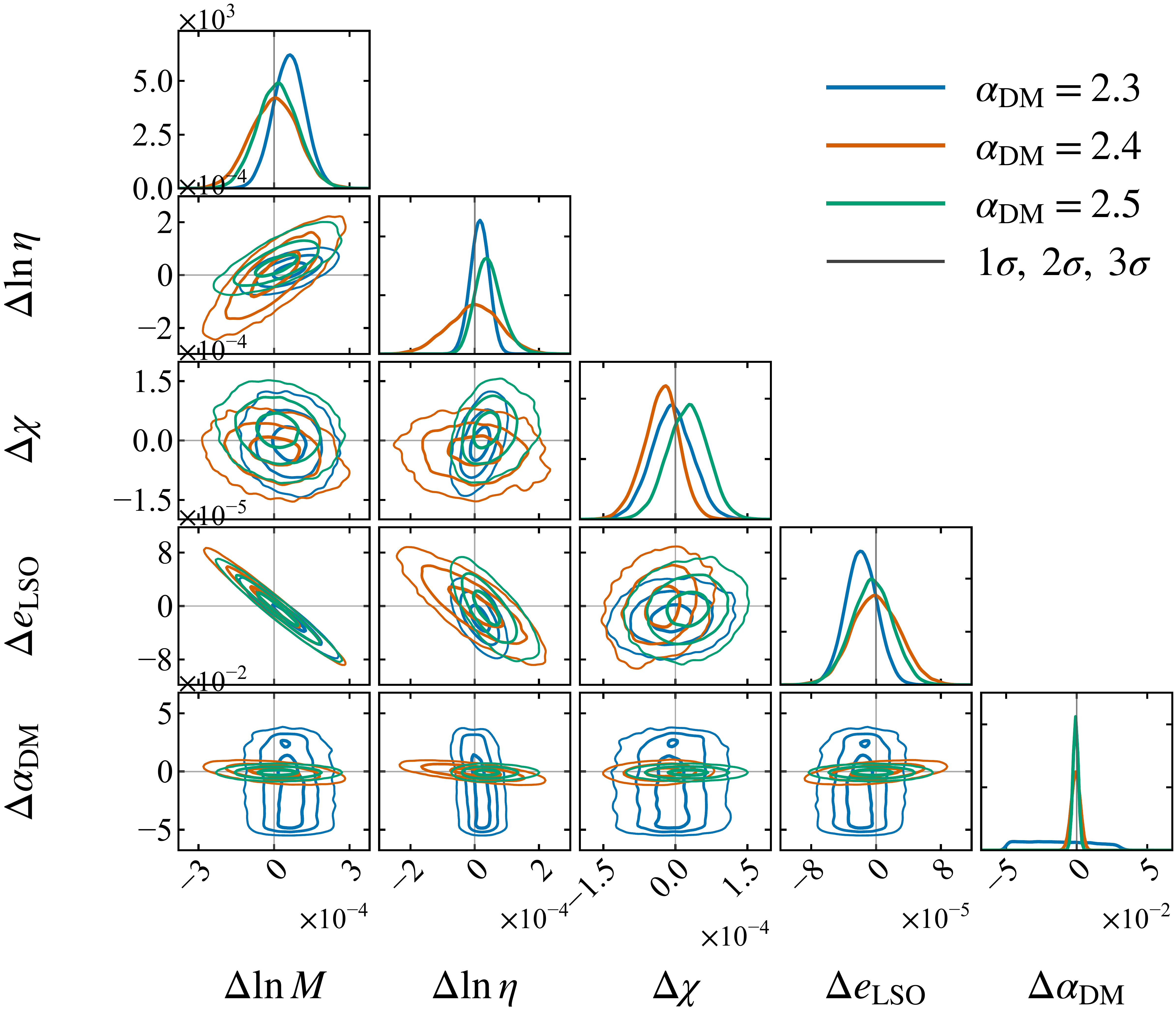}
\caption{Posterior distributions of five intrinsic parameters for three injected DM-spike slopes \(\alpha_{\rm DM}=2.3\), \(2.4\) and \(2.5\) are plotted.  The source parameters shown here are only  \(\{\Delta\ln M,\Delta\ln\eta,\Delta\chi,\Delta e_{\rm LSO}, \Delta\alpha_{\rm DM}\}\). The contours enclose the \(1\sigma\), \(2\sigma\) and \(3\sigma\) credible regions, and the  dashed reference lines mark zero offset from the injected values. The source parameter injected are keeping same with Fig.~\ref{fig:fim-mc-noise}.} \label{fig:alpha-comparison}
\end{figure}

To quantify the measurability of the DM-spike density profile, we perform
MCMC analyses for three representative values of the spike slope,
\(\alpha_{\rm DM}=2.3\), \(2.4\), and \(2.5\). For clarity, we display only
the parameters most relevant to the present discussion rather than the full
posterior parameter set. The corresponding marginalized posterior
distributions are shown in Fig.~\ref{fig:alpha-comparison}, where the blue,
orange, and green contours correspond to injections with
\(\alpha_{\rm DM}=2.3\), \(2.4\), and \(2.5\), respectively. The horizontal
and vertical dashed lines indicate the injected values. For the case of \(\alpha_{\rm DM}=2.3\), the bound on the intrinsic parameters  are list as follows
\begin{align}
\Delta \ln M
&= \left(6.02^{+6.23}_{-6.33}\right)\times10^{-5},
\nonumber\\
\Delta \ln \eta
&= \left(1.57^{+2.71}_{-2.82}\right)\times10^{-5},
\nonumber\\
\Delta\chi
&= \left(-0.77^{+4.30}_{-4.23}\right)\times10^{-5},
\nonumber\\
\Delta e_{\rm LSO}
&= \left(-1.83^{+1.94}_{-1.91}\right)\times10^{-5},
\nonumber\\
\Delta\alpha_{\rm DM}
&= \left(-1.31^{+2.78}_{-2.54}\right)\times10^{-2}.
\nonumber
\end{align}

For the case of \(\alpha_{\rm DM}=2.4\), we obtain
\begin{align}
\Delta \ln M
&= \left(0.16^{+9.10}_{-9.60}\right)\times10^{-5},
\nonumber\\
\Delta \ln \eta
&= \left(-0.04^{+7.66}_{-8.34}\right)\times10^{-6},
\nonumber\\
\Delta\chi
&= \left(-2.76^{+3.39}_{-3.91}\right)\times10^{-5},
\nonumber\\
\Delta e_{\rm LSO}
&= \left(-0.05^{+3.00}_{-2.84}\right)\times10^{-5},
\nonumber\\
\Delta\alpha_{\rm DM}
&= \left(-0.62^{+2.51}_{-2.58}\right)\times10^{-3}.
\nonumber
\end{align}
For the steepest spike considered here, \(\alpha_{\rm DM}=2.5\), the
corresponding offsets are
\begin{align}
\Delta \ln M
&= \left(1.46^{+8.19}_{-8.03}\right)\times10^{-5},
\nonumber\\
\Delta \ln \eta
&= \left(4.09^{+4.38}_{-3.76}\right)\times10^{-5},
\nonumber\\
\Delta\chi
&= \left(2.61^{+4.08}_{-4.28}\right)\times10^{-5},
\nonumber\\
\Delta e_{\rm LSO}
&= \left(-0.50^{+2.47}_{-2.53}\right)\times10^{-5},
\nonumber\\
\Delta\alpha_{\rm DM}
&= \left(-5.89^{+4.35}_{-8.22}\right)\times10^{-5}.
\nonumber
\end{align}
A clear trend emerges as the spike becomes steeper: the posterior of
\(\alpha_{\rm DM}\) becomes more localized, the measurement error is improved. This behavior reflects the increasing environmental imprint
produced by a steeper DM distribution. In particular, the uncertainty in \(\alpha_{\rm DM}\) decreases from
\(\mathcal{O}(10^{-2})\) for \(\alpha_{\rm DM}=2.3\), to
\(\mathcal{O}(10^{-3})\) for \(\alpha_{\rm DM}=2.4\), and further to
\(\mathcal{O}(10^{-5})\) for \(\alpha_{\rm DM}=2.5\). Thus, over the range
considered here, the measurability of the DM-spike slope improves by more
than two orders of magnitude, while the measurement precision of the
intrinsic EMRI parameters is affected only mildly. This indicates that the
stronger environmental modification associated with a steeper spike is
encoded primarily in the determination of \(\alpha_{\rm DM}\), rather than
being absorbed into large shifts of the standard source parameters.

To understand the parameter degeneracies underlying the Bayesian inference,
we examine the marginalized contours in Fig.~\ref{fig:alpha-comparison}.
Several pronounced correlations are evident among the measurement error of intrinsic EMRI parameters. In particular, the central BH mass $\Delta\ln M$ and $\Delta\ln\eta$ exhibit strong positive
correlations. This behavior is expected because these parameters enter the
secular orbital evolution and accumulated GW phase in a coupled manner. By contrast, \(e_{\rm LSO}\) is anti-correlated with \(M\) and \(\eta\). It may be due to that, the pronounced anti-correlation between \(M\) and \(e_{\rm LSO}\) can be
understood more directly from the termination frequency adopted in AK
model. The orbital frequency at the LSO is approximately
given by
\begin{equation}
    \nu_{\rm LSO}
    =
    \frac{c^3}{2\pi G M}
    \left(
    \frac{1-e_{\rm LSO}^{2}}
         {6+2e_{\rm LSO}}
    \right)^{3/2}.
\label{eq:lso_frequency_eccentricity}
\end{equation}
At a fixed value of \(\nu_{\rm LSO}\), the quantities \(M\) and
\(e_{\rm LSO}\) cannot vary independently. In particular, increasing \(M\)
reduces the prefactor \(1/M\); this change must be compensated by an increase
of the eccentricity-dependent factor in Eq.~\eqref{eq:lso_frequency_eccentricity}.
Since this factor decreases monotonically with \(e_{\rm LSO}\), the
compensation requires a smaller value of \(e_{\rm LSO}\). This directly
explains the negative \(M\)--\(e_{\rm LSO}\) correlation observed in the
posterior.  Consequently, correlated variations of \(M\), \(\eta\) and \(e_{\rm LSO}\) can give rise to the additional
anti-correlations seen in Fig.~\ref{fig:alpha-comparison}.

\begin{figure}[!t]
\centering
\includegraphics[width=0.98\linewidth,height=0.78\textheight,keepaspectratio]{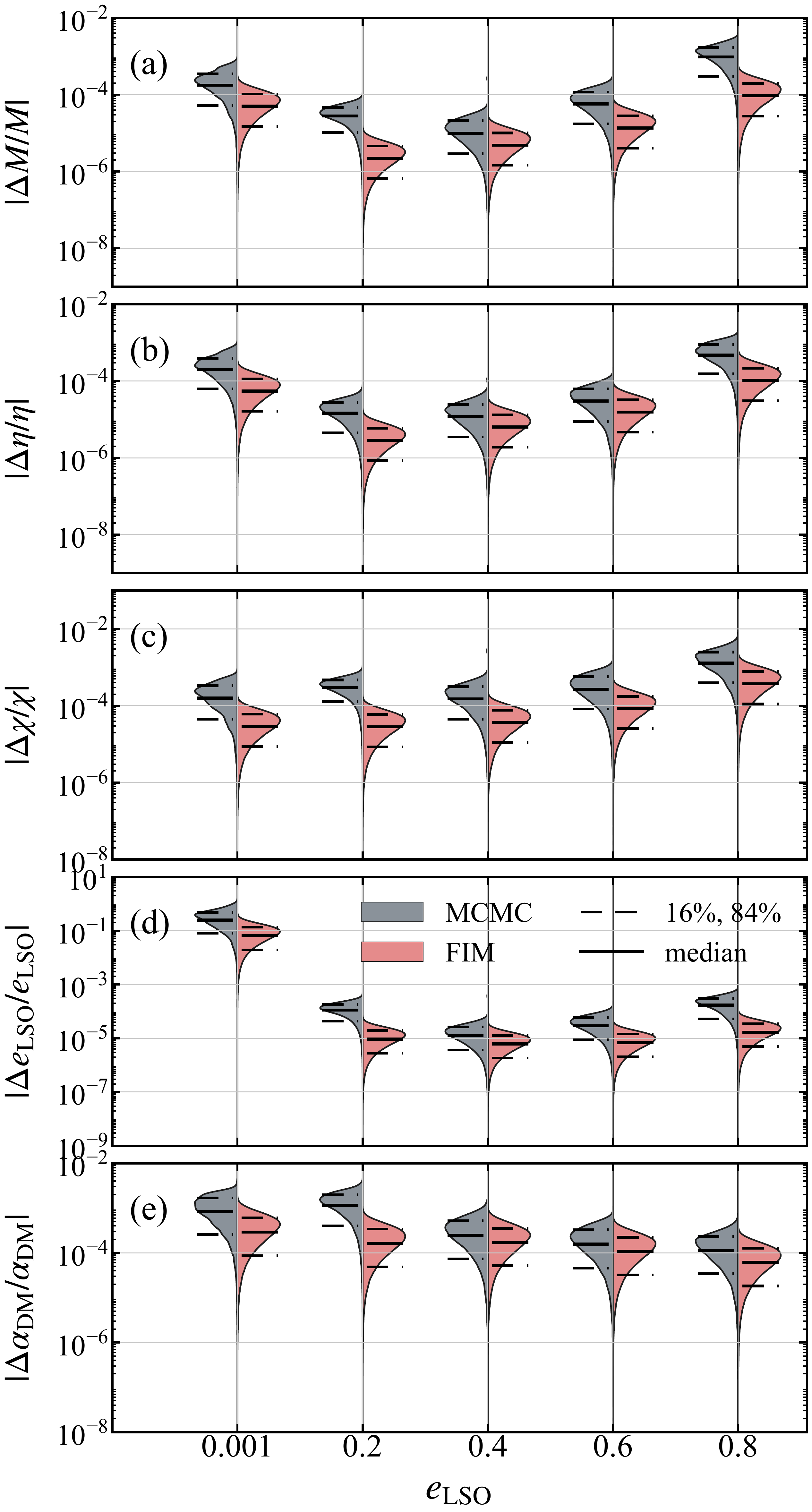}
\caption{Constraints of intrinsic parameters from MCMC and FIM analyses are shown, considering \(e_{\rm LSO}\in\{0.001,0.2,0.4,0.6,0.8\}\). The gray and red half-violins denote the MCMC and FIM error distributions, respectively. The five panels show the posterior
distributions of the absolute relative errors in \(M\), \(\eta\), \(\chi\),
\(e_{\rm LSO}\), and \(\alpha_{\rm DM}\). 
The solid horizontal bars
indicate the medians of the distributions, and the dashed bars represent the \(16\%\) and \(84\%\) confidence bounds. Note that the other source parameter injected are keeping same with Fig.~\ref{fig:fim-mc-noise}. }\label{fig:error-eLSO}
\end{figure}

\subsection{Impact of orbital plunge eccentricity on parameter estimation}
To investigate the role of orbital plunge eccentricity for constraining the source parameters with two statistical methods, we adopt the FIM and MCMC methods to plot ~\autoref{fig:error-eLSO} showing the relative parameter uncertainties as functions of the eccentricity \(e_{\rm LSO}\) of LSO. From top to bottom,
the five panels correspond to the relative uncertainties in the panel of
(a) \( M\), (b) \(\eta\), (c) \(\chi\), (d) \(e_{\rm LSO}\) and (e) \(\alpha_{\rm DM}\), respectively.
The gray distributions are obtained from the MCMC posterior samples, while the red distributions are obtained from the  FIM posterior samples. The solid horizontal bars indicate the median values of the corresponding distributions,
and the dashed bars mark the associated \(16\%\) and \(84\%\) confidence bounds. It is clear that the FIM error widths are smaller than those from MCMC. 
Over most of the eccentricity range considered here, the characteristic uncertainties inferred from the MCMC
and Fisher analyses are broadly consistent. The MCMC
results, however, exhibit asymmetric and non-Gaussian
tails that become more prominent for some of the high-
eccentricity configurations. Such features are not cap-
tured by the local Gaussian approximation underlying
the Fisher analysis.
The uncertainties in the intrinsic parameters \(M\), \(\eta\) and
\(\chi\) show a nonmonotonic dependence on \(e_{\rm LSO}\).
The constraints generally improve as the orbit becomes moderately
eccentric, but deteriorate again for the largest eccentricity considered,
$e_{\rm LSO}=0.8$. A plausible interpretation is that the AK model can fail to capture the full relativistic effect of more eccentric EMRI orbits. Moderate eccentricity enriches the harmonic content of the
waveform and provides additional information that can help break parameter
degeneracies. At very large eccentricity, however, there is a larger error in the AK waveform may occur. For the moderate and smaller eccentricities, the AK model can reflect the main features of an EMRI system in complex astrophysical environments, and our result can be a preliminary test of constraints on DM spikes.

The eccentricity uncertainty itself exhibits a somewhat different trend, as shown in panel (d). For the nearly circular configuration,
\(e_{\rm LSO}=0.001\), the fractional uncertainty with LISA becomes large. This reflects both the weak sensitivity of the waveform to a very small residual eccentricity and the fact that the uncertainty is normalized by
a very small injected value. At intermediate eccentricities, the eccentricity leaves a more pronounced imprint on the harmonic structure and frequency evolution of the AK waveform, leading to the tighter fractional constraints. The uncertainty increases again for \(e_{\rm LSO}=0.8\), possibly because the AK model fails to describe the richer harmonic content.
As shown in panel (e), the relative uncertainty in \(\alpha_{\rm DM}\) 
displays a comparable order of magnitude for the low- and intermediate-eccentricity configurations.
These source-data values show that the FIM gives smaller local uncertainties than the full MCMC posteriors, while preserving the qualitative eccentricity dependence.
Overall, Fig.~\ref{fig:error-eLSO} indicates that moderately eccentric EMRIs provide the rigorous bound on the DM-spike profile using the AK model.

\subsection{Role of eccentric harmonics in breaking parameter degeneracies}
\label{Sec:Role of eccentric harmonics}

\begin{figure}[!t]
\centering
\includegraphics[width=0.98\linewidth,height=0.78\textheight,keepaspectratio]{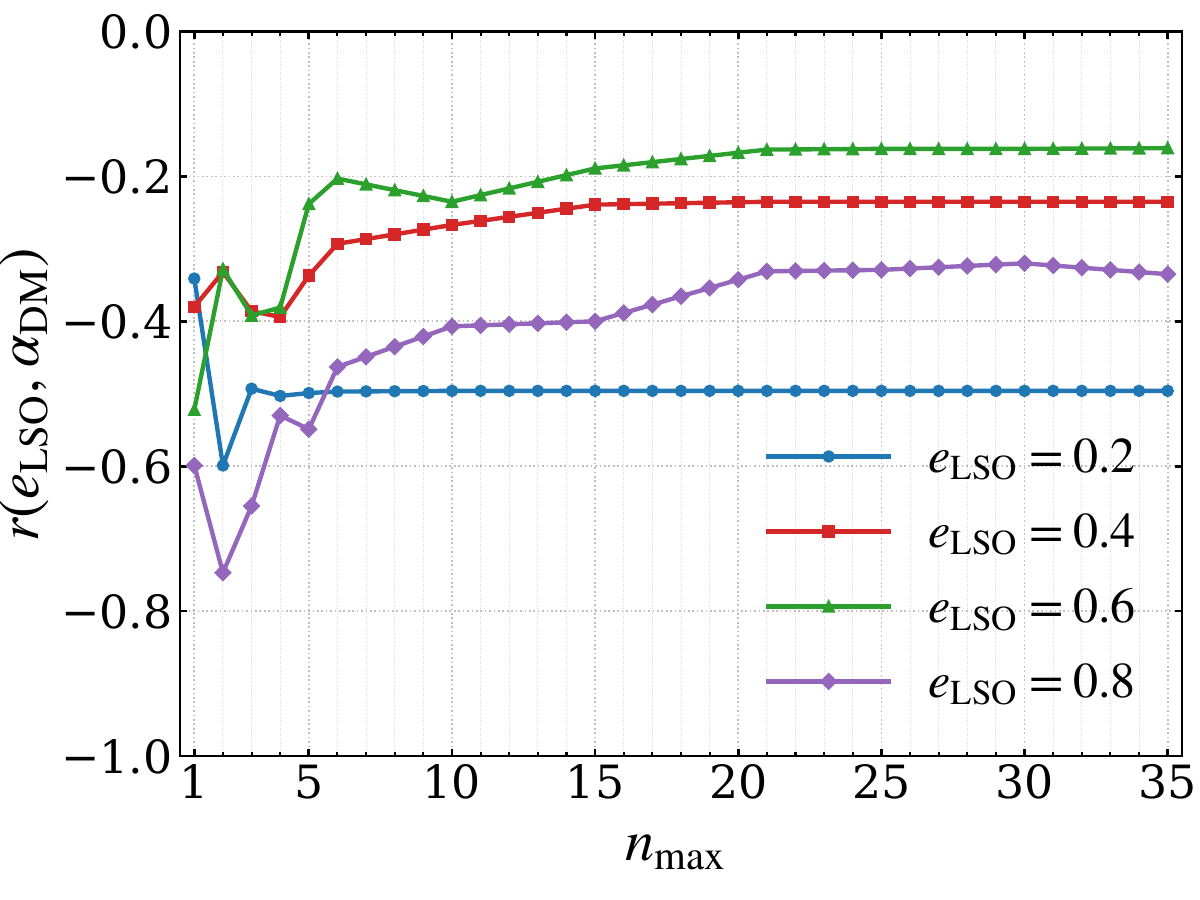}
\caption{Correlation coefficient in FIM analysis between \(e_{\rm LSO}\) and
\(\alpha_{\rm DM}\) as a function of the maximum harmonic order
\(n_{\max}\).  The four curves correspond to representative terminal
eccentricities \(e_{\rm LSO}=0.2,0.4,0.6\) and \(0.8\).  The non-monotonic
behavior shows that eccentric harmonics can either strengthen or weaken the
degeneracy, depending on the eccentricity and on the harmonic content retained
in the waveform. At \(n_{\max}=21\), the corrected source-data values are \(r(e_{\rm LSO},\alpha_{\rm DM})=\{-0.496,-0.235,-0.163,-0.331\}\).
 The other source parameter injected are keeping same with Fig.~\ref{fig:fim-mc-noise}.
}
\label{fig:r_e_alpha_nmax}
\end{figure}

To understand how the eccentric orbital harmonics affect the degeneracy between the plunge eccentricity \(e_{\rm LSO}\) and the DM-spike slope
\(\alpha_{\rm DM}\), we calculate the correlation coefficient
\(r(e_{\rm LSO},\alpha_{\rm DM})\) as a function of the maximum harmonic
order \(n_{\max}\), as shown in Fig.~\ref{fig:r_e_alpha_nmax}.

For \(e_{\rm LSO}=0.2\), the magnitude of the correlation decreases as higher harmonics of orbits are included and becomes stable after the fifth harmonic, with \(r\simeq-0.496\). For \(e_{\rm LSO}=0.4\), this effect is more evident: the initially strong negative correlation stabilizes after higher harmonics are included (\(n_{\max}=15\)), and the correlation coefficient is smaller than that for \(e_{\rm LSO}=0.2\). For the larger eccentricity \(e_{\rm LSO}=0.6\), higher-order harmonics (\(n_{\max}=20\)) are required, and the correlation coefficient is further reduced to about \(-0.16\). This indicates that more harmonics are included at moderate eccentricity, helping to break the degeneracy between \(e_{\rm LSO}\) and \(\alpha_{\rm DM}\). For \(e_{\rm LSO}=0.8\), the correlation depending on \(n_{\max}\) becomes irregular; for example, it reaches \(r=-0.747\) at \(n_{\max}=2\), then approaches \(r=-0.331\) at \(n_{\max}=21\) and \(r=-0.335\) at \(n_{\max}=35\). It indicates that orbital dynamics in the highly eccentric regime should be
modeled with caution, because the FIM becomes increasingly ill
conditioned after using inaccurate waveform models, reducing the reliability of the inferred correlation structure.
The numerical diagnostics supporting this statement, including the SNR and the relative uncertainties for each \(n_{\max}\), are listed in
Appendix~\ref{app:harmonic_truncation}.

Overall, the effectiveness of higher harmonics in breaking the
\(e_{\rm LSO}\)--\(\alpha_{\rm DM}\) degeneracy depends strongly on the
orbital eccentricity. For the configurations considered here, moderately
eccentric systems, particularly around \(e_{\rm LSO}\simeq0.4\sim0.6\), provide the
most favorable conditions for separating the eccentricity and DM effects.
This behavior also helps explain the nonmonotonic dependence of the
\(\alpha_{\rm DM}\) constraint on orbital eccentricity.
\begin{figure*}
\centering    \includegraphics[width=1\textwidth,height=1\textheight,keepaspectratio]{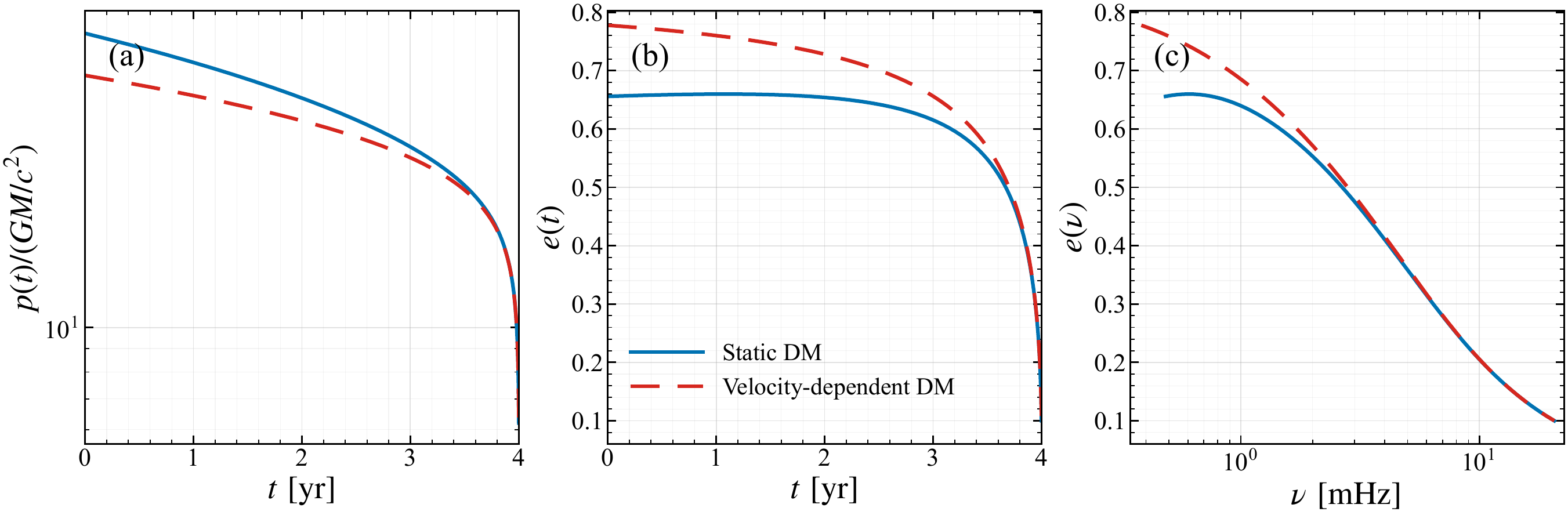}
 \caption{Orbital evolution for Static DM and Velocity-dependent DM, fixing the eccentricity at the LSO to \(e_{\rm LSO}=0.1\). Panel (a) shows the semilatus-rectum evolution, \(p(t)/(GM/c^2)\), panel (b) shows the eccentricity evolution \(e(t)\), and panel (c) shows the eccentricity as a function of orbital frequency, \(e(\nu)\). The remaining source parameters are
\(M=1.0\times10^{5}M_{\odot}\), \(\mu=10M_{\odot}\),
\(\chi=0.1\), and \(t_c=4~{\rm yr}\).
The terminal orbital frequency is determined by
\(\nu_{\rm LSO}=c^{3}(2\pi GM)^{-1}
[(1-e_{\rm LSO}^{2})/(6+2e_{\rm LSO})]^{3/2}\).}\label{fig:velocity-static-dynamic-orbit}
\end{figure*}

\subsection{DM model with a velocity distribution}
\label{Sec:velocity_distribution_dm_model}
If the DM particles in the DM-spike have a nontrivial velocity distribution, DF is not produced only by particles moving more slowly than the compact object; particles moving faster than the compact object can also modify the gravitational wake and contribute to the drag coefficient~\cite{Dosopoulou_2024,Zhou2025GeneralDF}. We therefore study how this effect modifies the orbital-evolution equations. This extension does not change the kinematic relations in Eqs.~\eqref{eq:A2_like}, \eqref{eq:v_decomposition_osculating}, \eqref{eq:orbital_average}, \eqref{eq:nu_p_e}, and \eqref{eq:avg_dnu_dphi}; it only replaces the constant drag coefficient by a phase-space dependent one. Following the correction of DM particles on orbital evolution with treatment of Chandrasekhar in Ref.~\cite{Dosopoulou_2024}, a normalized local distribution of DM particles \(f_{\rm DM}(r,u)\), one can obtain \(4\pi\int_0^{v_{\rm esc}}f_{\rm DM}(r,u)u^2du=1\) by defining
\begin{align}
\mathcal A(r,v)
&=4\pi\int_{0}^{v} f_{\rm DM}(r,u)u^2du,
\nonumber\\
\mathcal B(r,v)
&=4\pi\int_{v}^{v_{\rm esc}(r)}
f_{\rm DM}(r,u)u^2
\nonumber\\
&\quad\times
\ln\!\left(\frac{u+v}{u-v}\right)du,
\nonumber\\
\mathcal D(r,v)
&=-8\pi v\int_{v}^{v_{\rm esc}(r)}
f_{\rm DM}(r,u)u\,du,
\nonumber\\
\mathcal C_v(r,v)
&=I_\lambda+I_v\mathcal A(r,v)+\mathcal B(r,v)+\mathcal D(r,v),
\label{eq:velocity_drag_coefficient}
\end{align}
with DM particle's velocity $u$ and orbital velocity $v$. Here \(\mathcal A\) is the contribution from DM particles slower than the compact object, whereas \(\mathcal B+\mathcal D\) reflects the velocity of DM particles faster than the compact object. The velocity-weighted terms of DM contribution  influences on orbital orbit by the following equations
\begingroup\small
\begin{align}
\avdot{\nu}_{\rm DM}^{(v)}
&=
2\pi\nu^2
\frac{4\mu\rhosp\rsp^{\alpha_{\rm DM}}}{M^2}
\notag\\
&\quad\times
(GM)^{(3-\alpha_{\rm DM})/3}
\notag\\
&\quad\times
(1-e^2)^{3-\alpha_{\rm DM}}
(2\pi\nu)^{-2(3-\alpha_{\rm DM})/3}
\notag\\
&\quad\times
\left[
\frac{3}{2}\IDMzero^{(v)}
+\frac{3e}{1-e^2}\IDMone^{(v)}
\right],
\label{eq:velocity_dnu_dt}
\\
\avdot{e}_{\rm DM}^{(v)}
&=
-(2\pi\nu)
\frac{4\mu\rhosp\rsp^{\alpha_{\rm DM}}}{M^2}
\notag\\
&\quad\times
(GM)^{(3-\alpha_{\rm DM})/3}
\notag\\
&\quad\times
(1-e^2)^{3-\alpha_{\rm DM}}
(2\pi\nu)^{-2(3-\alpha_{\rm DM})/3}
\notag\\
&\quad\times
\IDMone^{(v)},
\label{eq:velocity_de_dt}
\\
\IDMzero^{(v)}
&=
\frac{1}{GM(1-e^2)^3}
\notag\\
&\quad\times
\int_0^{2\pi}
\mathcal C_v(\phi)
\mathcal P_v(\phi)^{-3/2}
\notag\\
&\quad\times
\mathcal Q_v(\phi)^3
\mathcal R_v(\phi)^{\alpha_{\rm DM}-2}\,d\phi,
\label{eq:velocity_IDMzero}
\\
\IDMone^{(v)}
&=
\frac{1}{GM(1-e^2)^3}
\notag\\
&\quad\times
\int_0^{2\pi}
\mathcal C_v(\phi)(e+\cos\phi)
\notag\\
&\quad\times
\mathcal P_v(\phi)^{-3/2}
\mathcal Q_v(\phi)^3
\mathcal R_v(\phi)^{\alpha_{\rm DM}-2}\,d\phi.
\label{eq:velocity_IDMone}
\end{align}
\endgroup
\begingroup\small
\begin{equation}
\begin{aligned}
\mathcal Q_v(\phi)
&=
(GM)^{1/3}(1-e^2)
-2R_s(2\pi\nu)^{2/3}(1+e\cos\phi),\\
\mathcal P_v(\phi)
&=1+2e\cos\phi+e^2,
\\
\mathcal R_v(\phi)
&=1+e\cos\phi.
\end{aligned}
\label{eq:velocity_QPR_definitions}
\end{equation}
\endgroup
From these equations, we plot the orbital parameters evolution under the influence of static and velocity-dependent DM in FIG.~\ref{fig:velocity-static-dynamic-orbit}, setting a fixed plunge orbital parameters. One can find that the  velocity-dependent DM can amplify the dynamical friction process between secondary compact and DM environments, leading to a smaller binary separation comparing to the static DM case. The effect of velocity-dependent DM particle increase the evolution of orbital eccentricity. We indeed observe that the velocity-dependent DM model can result in a deviation of orbital parameters visible. 

\begin{figure}
\centering    \includegraphics[width=0.49\textwidth,height=0.50\textheight,keepaspectratio]{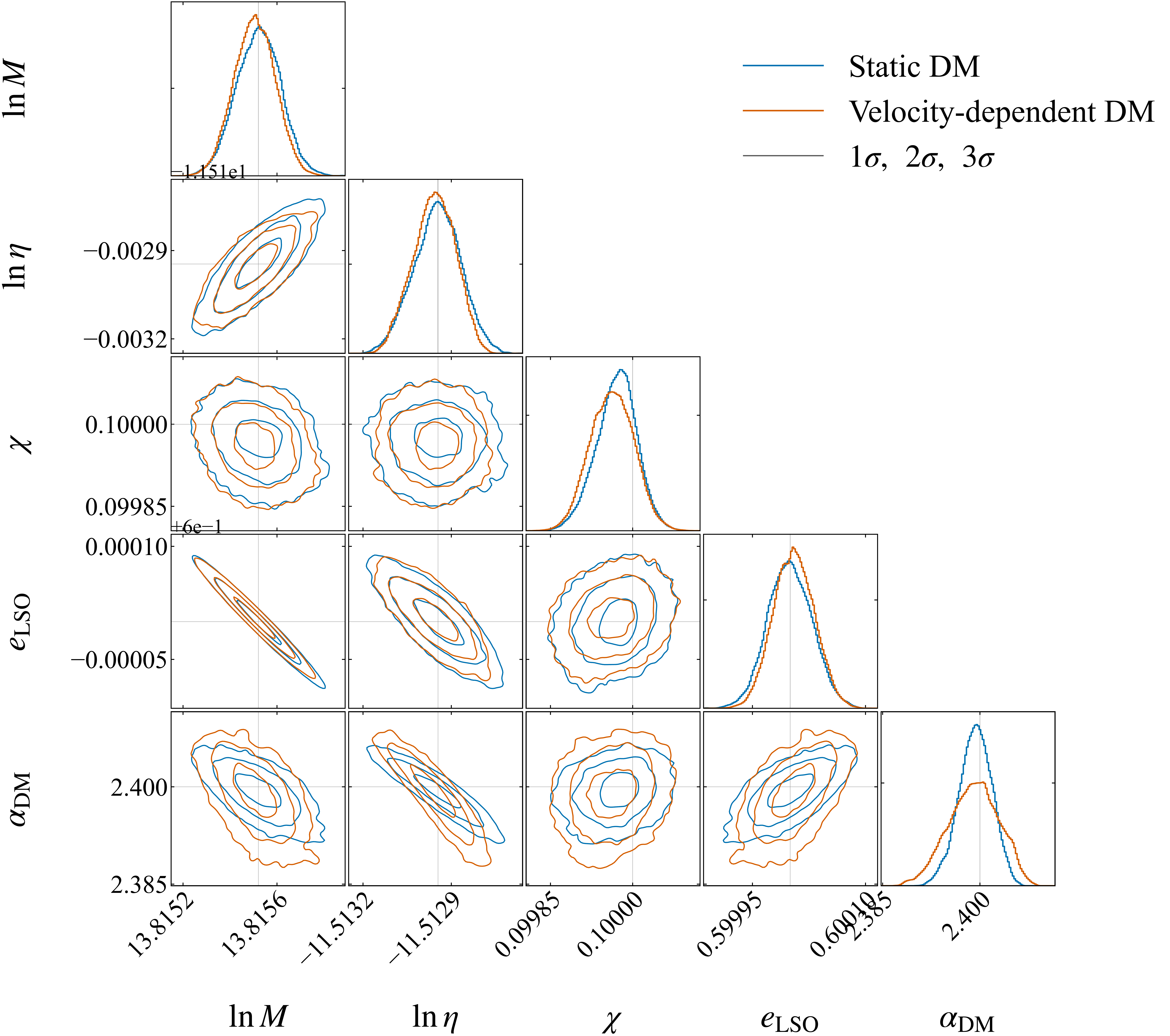}
 \caption{
 Posterior distributions of five intrinsic parameters  are plotted for the static DM particle (blue contours) and velocity-dependent (yellow contours) DM models. The gray reference lines  mark the injected parameters \(\{\ln M,\ln\eta,\chi,e_{\rm LSO},\alpha_{\rm DM}\}\). The nested  contours enclose the \(1\sigma\), \(2\sigma\), and \(3\sigma\)
 credible regions. The other source parameter injected are keeping same with Fig.~\ref{fig:fim-mc-noise}.}\label{fig:velocity-static-dynamic-corner}
\end{figure}

To study how the  velocity-dependent DM influences parameters estimation within MCMC analysis, we next present the posterior distributions in FIG~\ref{fig:velocity-static-dynamic-corner}, and compare the marginalized
posterior distributions of the intrinsic parameters obtained with the
static and velocity-dependent DM models using the modified AK waveform.
The injected parameters are 
\(M=1.0\times10^{6}M_{\odot}\), \(\mu=10M_{\odot}\),
\(\chi=0.1\), \(e_{\rm LSO}=0.6\), \(\alpha_{\rm DM} =2.4\) and \(t_c=1~{\rm yr}\).
The blue and yellow contours correspond to the static and
velocity-dependent DM models, respectively, with the nested contours
representing the \(1\sigma\), \(2\sigma\), and \(3\sigma\) credible regions.
The two models yield broadly consistent parameter recovery, although the
static DM model produces slightly narrower posterior distributions for some
parameters. This indicates that the velocity dependence of the DM
interaction introduces additional parameter degeneracies and mildly weakens
the constraints on the DM-dressed EMRI system.

\section{Conclusion}
\label{sec:conclusion}

In this work, we investigated the measurability of DM spikes
with eccentric EMRI gravitational waves. We incorporated the leading
dissipative effects of dynamical friction and accretion into the
AK waveform model and performed Bayesian parameter estimation
using MCMC sampling, supplemented by Fisher-matrix analyses.

The Bayesian analysis shows that the DM-induced modification becomes more
measurable as the spike becomes steeper. For the adopted spike prescription
and fiducial configurations, the characteristic uncertainty in
\(\alpha_{\rm DM}\) decreases from approximately
\(\mathcal{O}(10^{-2})\) for \(\alpha_{\rm DM}=2.3\) to
\(\mathcal{O}(10^{-5})\) for \(\alpha_{\rm DM}=2.5\). The Fisher analysis
provides a forecast estimate of the parameter uncertainties at high
SNR, while the MCMC posteriors reveal non-Gaussian structures and nonlinear
parameter correlations that are not captured by the Gaussian approximation.

A central result of this work is that the accuracy of DM inference does not
improve monotonically with orbital eccentricity. Moderately eccentric systems
provide the most favorable parameter recovery among the configurations
considered here. Around \(e_{\rm LSO}\simeq0.5\), the inclusion of higher
harmonics substantially reduces the correlation between \(e_{\rm LSO}\) and
\(\alpha_{\rm DM}\), helping to disentangle eccentric orbital dynamics from
DM-induced secular evolution. At larger eccentricities, this degeneracy can
persist, while for \(e_{\rm LSO}=0.8\) the accuracy of the AK description
deteriorates and the corresponding parameter constraints become less robust.
We further compare the MCMC posteriors obtained with static and
velocity-dependent DM prescriptions. The two models yield broadly consistent
parameter recovery, although the static-DM model gives slightly tighter
constraints for the configurations considered here, indicating that the
velocity dependence introduces additional parameter degeneracies.

The present results should be regarded as a proof-of-principle study.
The AK waveform does not provide the accuracy required for precision LISA
inference, and our environmental model neglects conservative effects of the
DM gravitational potential as well as the dynamical response and depletion
of the DM spike. Incorporating these effects into more accurate relativistic
EMRI waveform models will be important for assessing waveform systematics.
Nevertheless, our results show that the interplay between eccentric
harmonics and environmental dissipation provides a promising avenue for
probing dense DM distributions around massive black holes with future
space-based gravitational-wave observations.

\section{Acknowledgements}
We thank Huai-Ke Guo for helpful discussions. This work is supported by the National Natural Science Foundation of China under Grants No. 12347140 and No. 12405059, the Key Program of the Natural Science Foundation of Jiangxi Province under Grant No. 20232ACB201008, the Ganpo High-Level Innovative Talent Program, the Startup Fund for Advanced Talents of Putian University  Grant No. 2023141, Scientific Research (on Science and Technology) Projects for Young and Middle-Aged Teachers in Fujian province  Grant No. JZ240055, and the Natural Science Foundation of Fujian Province  Grant No. 2025J011042.

\appendix
\section{Full-parameter posterior corner plots}\label{app:full_posterior}
\begin{figure*}[!p]
\centering    \includegraphics[width=0.98\textwidth,height=0.86\textheight,keepaspectratio]{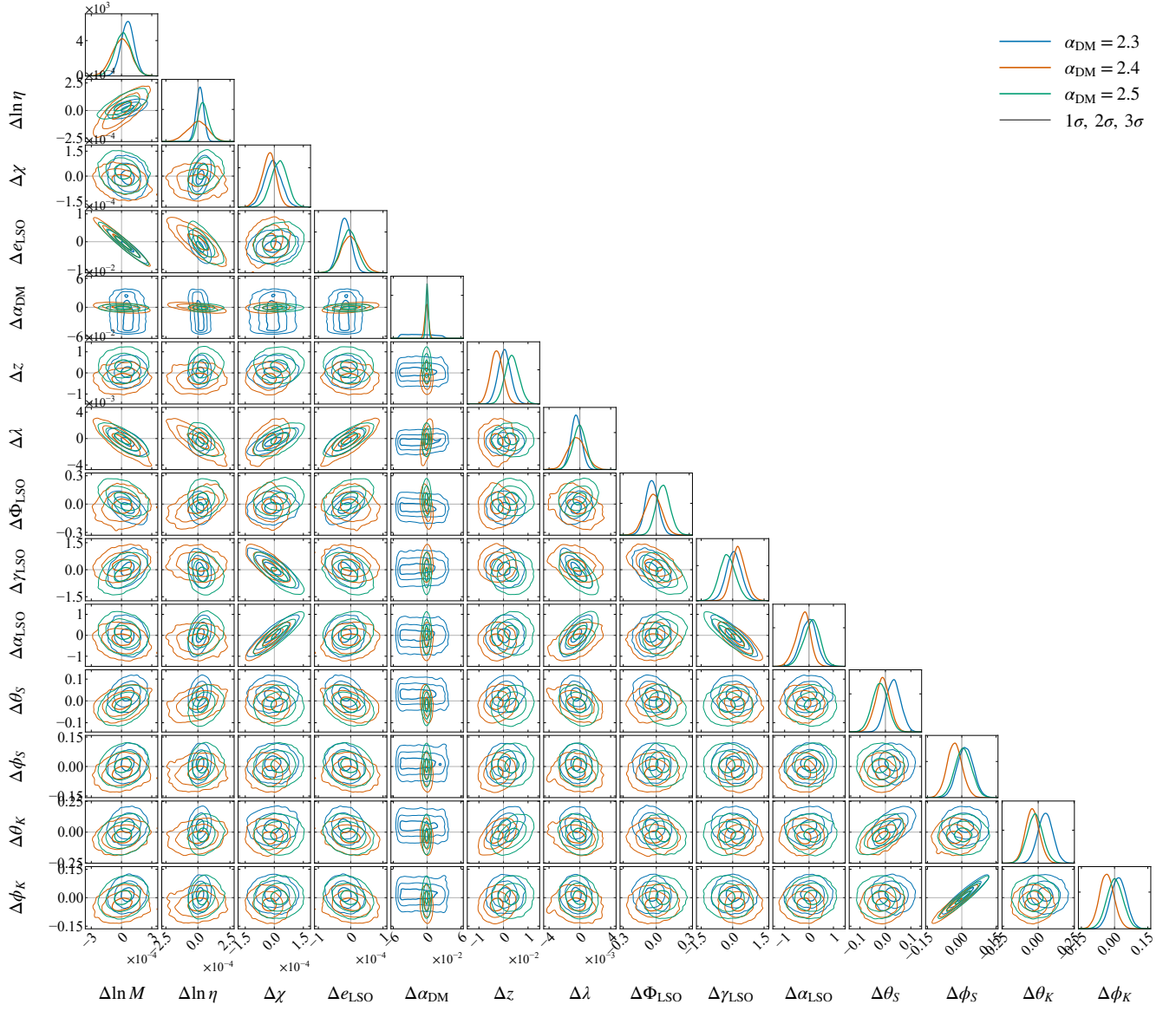}
 \caption{
Posterior distributions of 14 source parameters are plotted using GW modified by DM, considering three injected  DM-spike slopes \(\alpha_{\rm DM}=2.3\), \(2.4\), and \(2.5\). The nested  contours enclose the \(1\sigma\), \(2\sigma\), and \(3\sigma\) credible regions. In this MCMC analyses, the corner plot for full source parameters is corresponding to the intrinsic parameters in Fig.~\ref{fig:alpha-comparison},  the other source parameter injected are keeping same with Fig.~\ref{fig:fim-mc-noise}.}\label{fig:alpha-comparison-full14}
\end{figure*}

\begin{figure*}[!p]
\centering
\includegraphics[width=0.98\textwidth,height=0.86\textheight,keepaspectratio]{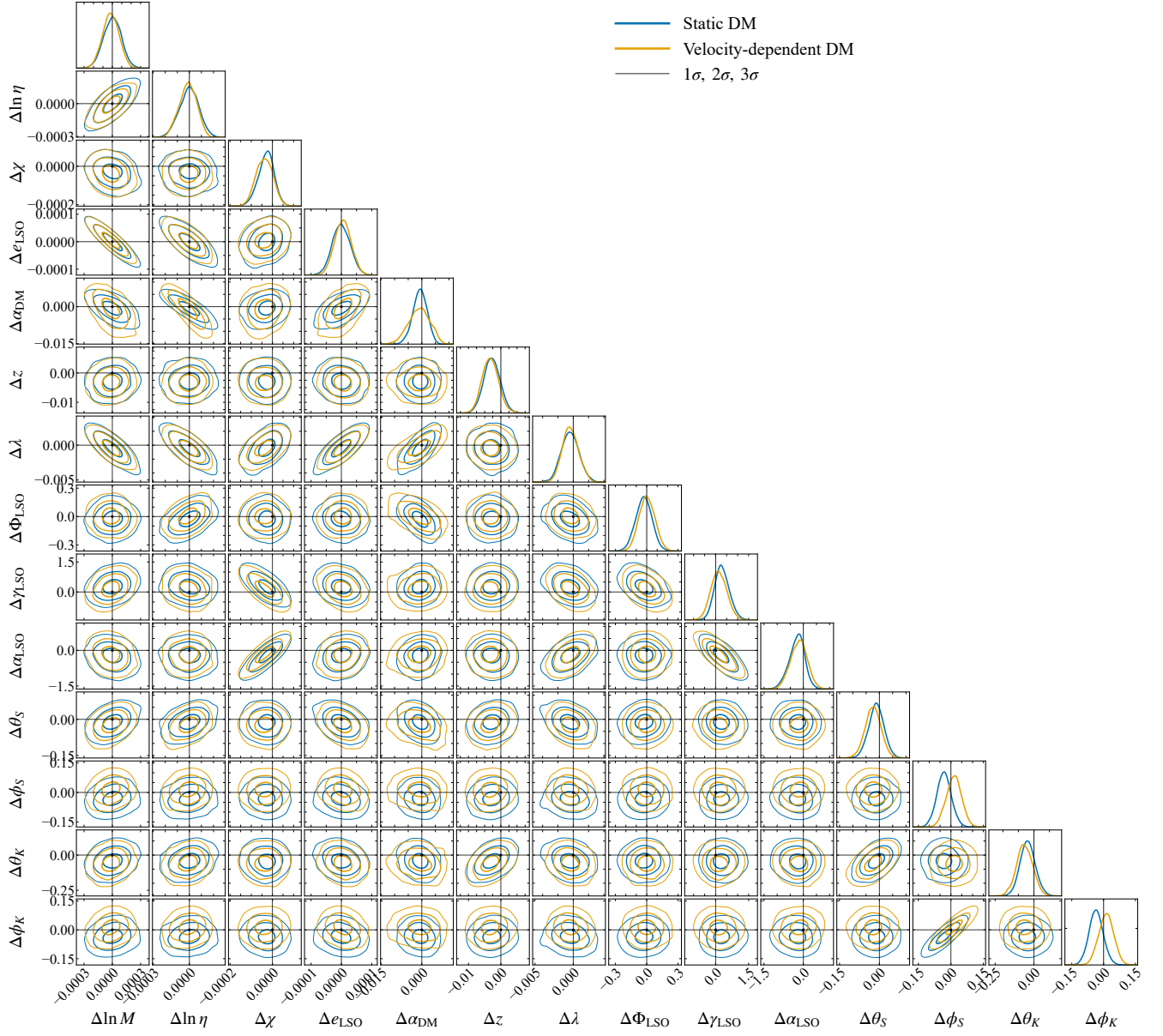}
\caption{
Posterior distributions of 14 source parameters are plotted for the static (blue contours) and velocity-dependent (yellow contours)  DM particles. The nested  contours enclose the \(1\sigma\), \(2\sigma\), and \(3\sigma\) credible regions.  In this MCMC analyses, the corner plot for full source parameters complements the intrinsic parameters in Fig.~\ref{fig:velocity-static-dynamic-corner}, the other source parameter injected are keeping same with Fig.~\ref{fig:fim-mc-noise}.}\label{fig:velocity-static-dynamic-full14}
\end{figure*}

In this appendix, we present the posterior distribution of 14 parameters in Eq.~\eqref{sources:emri}, 
the Beyesian analysis of full source parameter in Fig.~\ref{fig:alpha-comparison-full14} and Fig.~\ref{fig:velocity-static-dynamic-full14} can be severed as a supplement of  Beyesian analysis  of intrinsic parameter
in Fig.~\ref{fig:alpha-comparison} and Fig.~\ref{fig:velocity-static-dynamic-corner}. In two figures, the other parameters setting is same with the  corresponding figure.

\clearpage
\onecolumngrid
\section{Harmonic-truncation Fisher results}\label{app:harmonic_truncation}
In this appendix, we show the effect of harmonic-truncation on measurement error of parameter using FIM results, which is used to support the discussion in Sec.~\ref{Sec:Role of eccentric harmonics}.
The main text focuses on the changing trend of the parameters
\(e_{\rm LSO}\)--\(\alpha_{\rm DM}\) correlation, table~\ref{tab:harmonic_truncation_eLSO02}, \ref{tab:harmonic_truncation_eLSO04}, \ref{tab:harmonic_truncation_eLSO06} and \ccr{\ref{tab:harmonic_truncation_eLSO08}} outline the numerical values of correlation
$r(e_{\rm LSO}, \alpha_{\rm DM})$ and measurement error of intrinsic parameters for four cases of eccentricities $e\in\{0.2,0.4,0.6,0.8\}$, increasing the harmonic mode index $n_{\rm max}$. The injected parameters are the same as those listed in Table~\ref{tab:mcmc_priors_full} and we use only the five intrinsic parameters to construct the FIM.
One can observe that the measurement ability of each intrinsic parameter is 
imporved.

\input{appendix_harmonic_truncation_tables_core5_corrected_snr}

\clearpage
\twocolumngrid
\bibliographystyle{apsrev4-2}
\bibliography{reference}

\end{document}

%% file: appendix_harmonic_truncation_tables_core5_corrected_snr.tex
\begin{table}[!htbp]
\centering
\caption{Influence of maximum values of harmonic mode index $n_{\rm max}$
on the relative uncertainties of intrinsic parameter  in FIM analysis for the case of $e_{\rm LSO}=0.2$ is listed. The table outlines the relative parameter uncertainties, the correlation coefficient
between $e_{\rm LSO}$ and $\alpha_{\rm DM}$ and the corresponding SNR.
The varied parameters are $\ln M$, $\ln\eta$, $\chi$, $e_{\rm LSO}$, and $\alpha_{\rm DM}$. The other source parameter injected are keeping same with Fig.~\ref{fig:fim-mc-noise}.}\label{tab:harmonic_truncation_eLSO02}
\scriptsize
\renewcommand{\arraystretch}{1.25}
\setlength{\tabcolsep}{5.5pt}
\begin{tabular*}{\textwidth}{@{\extracolsep{\fill}} cccccccc}
\hline\hline
$n_{\max}$ & $\sigma_M/M$ & $\sigma_\eta/\eta$ & $\sigma_\chi/\chi$ & $\sigma_{e_{\rm LSO}}/e_{\rm LSO}$ & $\sigma_{\alpha_{\rm DM}}/\alpha_{\rm DM}$ & $r_{e\alpha}$ & SNR \\
\hline
1  & $3.73{\times}10^{-5}$ & $4.57{\times}10^{-5}$ & $2.39{\times}10^{-4}$ & $1.35{\times}10^{-4}$ & $2.24{\times}10^{-3}$ & $-0.341$ & 8.86 \\
2  & $5.41{\times}10^{-6}$ & $5.63{\times}10^{-6}$ & $5.01{\times}10^{-5}$ & $2.14{\times}10^{-5}$ & $3.09{\times}10^{-4}$ & $-0.599$ & 53.73 \\
3  & $3.33{\times}10^{-6}$ & $4.11{\times}10^{-6}$ & $4.15{\times}10^{-5}$ & $1.41{\times}10^{-5}$ & $2.15{\times}10^{-4}$ & $-0.493$ & 62.52 \\
4  & $3.11{\times}10^{-6}$ & $3.84{\times}10^{-6}$ & $4.00{\times}10^{-5}$ & $1.32{\times}10^{-5}$ & $2.03{\times}10^{-4}$ & $-0.503$ & 63.56 \\
5  & $3.08{\times}10^{-6}$ & $3.81{\times}10^{-6}$ & $3.98{\times}10^{-5}$ & $1.31{\times}10^{-5}$ & $2.01{\times}10^{-4}$ & $-0.499$ & 63.65 \\
6  & $3.06{\times}10^{-6}$ & $3.81{\times}10^{-6}$ & $3.97{\times}10^{-5}$ & $1.31{\times}10^{-5}$ & $2.01{\times}10^{-4}$ & $-0.497$ & 63.65 \\
10 & $3.05{\times}10^{-6}$ & $3.80{\times}10^{-6}$ & $3.96{\times}10^{-5}$ & $1.30{\times}10^{-5}$ & $2.01{\times}10^{-4}$ & $-0.496$ & 63.65 \\
15 & $3.05{\times}10^{-6}$ & $3.80{\times}10^{-6}$ & $3.96{\times}10^{-5}$ & $1.30{\times}10^{-5}$ & $2.01{\times}10^{-4}$ & $-0.496$ & 63.65 \\
21 & $3.05{\times}10^{-6}$ & $3.80{\times}10^{-6}$ & $3.96{\times}10^{-5}$ & $1.30{\times}10^{-5}$ & $2.01{\times}10^{-4}$ & $-0.496$ & 63.65 \\
25 & $3.05{\times}10^{-6}$ & $3.80{\times}10^{-6}$ & $3.96{\times}10^{-5}$ & $1.30{\times}10^{-5}$ & $2.01{\times}10^{-4}$ & $-0.496$ & 63.65 \\
30 & $3.05{\times}10^{-6}$ & $3.80{\times}10^{-6}$ & $3.96{\times}10^{-5}$ & $1.30{\times}10^{-5}$ & $2.01{\times}10^{-4}$ & $-0.496$ & 63.65 \\
35 & $3.05{\times}10^{-6}$ & $3.80{\times}10^{-6}$ & $3.96{\times}10^{-5}$ & $1.30{\times}10^{-5}$ & $2.01{\times}10^{-4}$ & $-0.496$ & 63.65 \\
\hline\hline
\end{tabular*}
\end{table}

\begin{table}[!htbp]
\centering
\caption{
Influence of maximum values of harmonic mode index $n_{\rm max}$
on the relative uncertainties of intrinsic parameter  in FIM analysis for the case of $e_{\rm LSO}=0.4$ is listed. The table outlines the relative parameter uncertainties, the correlation coefficient
between $e_{\rm LSO}$ and $\alpha_{\rm DM}$ and the corresponding SNR.
The varied parameters are $\ln M$, $\ln\eta$, $\chi$, $e_{\rm LSO}$, and $\alpha_{\rm DM}$. The other source parameter injected are keeping same with Fig.~\ref{fig:fim-mc-noise}.
}\label{tab:harmonic_truncation_eLSO04}
\scriptsize
\renewcommand{\arraystretch}{1.25}
\setlength{\tabcolsep}{5.5pt}
\begin{tabular*}{\textwidth}{@{\extracolsep{\fill}} cccccccc}
\hline\hline
$n_{\max}$ & $\sigma_M/M$ & $\sigma_\eta/\eta$ & $\sigma_\chi/\chi$ & $\sigma_{e_{\rm LSO}}/e_{\rm LSO}$ & $\sigma_{\alpha_{\rm DM}}/\alpha_{\rm DM}$ & $r_{e\alpha}$ & SNR \\
\hline
1  & $1.40{\times}10^{-4}$ & $1.53{\times}10^{-4}$ & $4.74{\times}10^{-4}$ & $1.55{\times}10^{-4}$ & $3.69{\times}10^{-3}$ & $-0.380$ & 8.94 \\
2  & $1.97{\times}10^{-5}$ & $3.95{\times}10^{-5}$ & $1.70{\times}10^{-4}$ & $2.42{\times}10^{-5}$ & $9.31{\times}10^{-4}$ & $-0.332$ & 23.33 \\
3  & $1.08{\times}10^{-5}$ & $1.96{\times}10^{-5}$ & $9.40{\times}10^{-5}$ & $1.36{\times}10^{-5}$ & $4.71{\times}10^{-4}$ & $-0.386$ & 39.47 \\
4  & $8.11{\times}10^{-6}$ & $1.22{\times}10^{-5}$ & $7.05{\times}10^{-5}$ & $1.03{\times}10^{-5}$ & $2.98{\times}10^{-4}$ & $-0.394$ & 48.03 \\
5  & $7.43{\times}10^{-6}$ & $1.05{\times}10^{-5}$ & $6.58{\times}10^{-5}$ & $9.63{\times}10^{-6}$ & $2.46{\times}10^{-4}$ & $-0.337$ & 50.89 \\
6  & $7.04{\times}10^{-6}$ & $9.68{\times}10^{-6}$ & $6.33{\times}10^{-5}$ & $9.21{\times}10^{-6}$ & $2.20{\times}10^{-4}$ & $-0.293$ & 51.64 \\
10 & $6.23{\times}10^{-6}$ & $8.90{\times}10^{-6}$ & $5.51{\times}10^{-5}$ & $8.18{\times}10^{-6}$ & $1.99{\times}10^{-4}$ & $-0.267$ & 51.86 \\
15 & $6.05{\times}10^{-6}$ & $8.83{\times}10^{-6}$ & $5.32{\times}10^{-5}$ & $7.96{\times}10^{-6}$ & $1.95{\times}10^{-4}$ & $-0.239$ & 51.90 \\
21 & $6.02{\times}10^{-6}$ & $8.82{\times}10^{-6}$ & $5.29{\times}10^{-5}$ & $7.92{\times}10^{-6}$ & $1.94{\times}10^{-4}$ & $-0.235$ & 51.90 \\
25 & $6.02{\times}10^{-6}$ & $8.82{\times}10^{-6}$ & $5.29{\times}10^{-5}$ & $7.92{\times}10^{-6}$ & $1.94{\times}10^{-4}$ & $-0.235$ & 51.90 \\
30 & $6.02{\times}10^{-6}$ & $8.82{\times}10^{-6}$ & $5.29{\times}10^{-5}$ & $7.92{\times}10^{-6}$ & $1.94{\times}10^{-4}$ & $-0.235$ & 51.90 \\
35 & $6.02{\times}10^{-6}$ & $8.82{\times}10^{-6}$ & $5.29{\times}10^{-5}$ & $7.92{\times}10^{-6}$ & $1.94{\times}10^{-4}$ & $-0.235$ & 51.90 \\
\hline\hline
\end{tabular*}
\end{table}

\begin{table}[!htbp]
\centering
\caption{
Influence of maximum values of harmonic mode index $n_{\rm max}$
on the relative uncertainties of intrinsic parameter  in FIM analysis for the case of medium eccentricity $e_{\rm LSO}=0.6$ is listed. The table outlines the relative parameter uncertainties, the correlation coefficient
between $e_{\rm LSO}$ and $\alpha_{\rm DM}$ and the corresponding SNR.
The varied parameters are $\ln M$, $\ln\eta$, $\chi$, $e_{\rm LSO}$, and $\alpha_{\rm DM}$. The other source parameter injected are keeping same with Fig.~\ref{fig:fim-mc-noise}.
}\label{tab:harmonic_truncation_eLSO06}
\scriptsize
\renewcommand{\arraystretch}{1.25}
\setlength{\tabcolsep}{5.5pt}
\begin{tabular*}{\textwidth}{@{\extracolsep{\fill}} cccccccc}
\hline\hline
$n_{\max}$ & $\sigma_M/M$ & $\sigma_\eta/\eta$ & $\sigma_\chi/\chi$ & $\sigma_{e_{\rm LSO}}/e_{\rm LSO}$ & $\sigma_{\alpha_{\rm DM}}/\alpha_{\rm DM}$ & $r_{e\alpha}$ & SNR \\
\hline
1  & $2.93{\times}10^{-4}$ & $5.67{\times}10^{-4}$ & $8.59{\times}10^{-4}$ & $1.31{\times}10^{-4}$ & $3.36{\times}10^{-3}$ & $-0.522$ & 4.88 \\
2  & $1.25{\times}10^{-4}$ & $2.95{\times}10^{-4}$ & $6.80{\times}10^{-4}$ & $5.96{\times}10^{-5}$ & $1.80{\times}10^{-3}$ & $-0.328$ & 6.15 \\
3  & $5.81{\times}10^{-5}$ & $1.05{\times}10^{-4}$ & $3.40{\times}10^{-4}$ & $2.82{\times}10^{-5}$ & $6.49{\times}10^{-4}$ & $-0.392$ & 12.07 \\
4  & $3.52{\times}10^{-5}$ & $6.45{\times}10^{-5}$ & $2.24{\times}10^{-4}$ & $1.74{\times}10^{-5}$ & $4.02{\times}10^{-4}$ & $-0.381$ & 19.46 \\
5  & $2.59{\times}10^{-5}$ & $4.10{\times}10^{-5}$ & $1.76{\times}10^{-4}$ & $1.31{\times}10^{-5}$ & $2.55{\times}10^{-4}$ & $-0.238$ & 26.74 \\
6  & $2.08{\times}10^{-5}$ & $2.88{\times}10^{-5}$ & $1.41{\times}10^{-4}$ & $1.05{\times}10^{-5}$ & $1.77{\times}10^{-4}$ & $-0.203$ & 32.76 \\
10 & $1.67{\times}10^{-5}$ & $2.14{\times}10^{-5}$ & $1.13{\times}10^{-4}$ & $8.54{\times}10^{-6}$ & $1.32{\times}10^{-4}$ & $-0.235$ & 39.22 \\
15 & $1.64{\times}10^{-5}$ & $2.07{\times}10^{-5}$ & $1.13{\times}10^{-4}$ & $8.38{\times}10^{-6}$ & $1.24{\times}10^{-4}$ & $-0.189$ & 39.42 \\
21 & $1.57{\times}10^{-5}$ & $2.03{\times}10^{-5}$ & $1.10{\times}10^{-4}$ & $8.08{\times}10^{-6}$ & $1.20{\times}10^{-4}$ & $-0.163$ & 39.60 \\
25 & $1.56{\times}10^{-5}$ & $2.02{\times}10^{-5}$ & $1.09{\times}10^{-4}$ & $8.01{\times}10^{-6}$ & $1.19{\times}10^{-4}$ & $-0.162$ & 39.64 \\
30 & $1.56{\times}10^{-5}$ & $2.02{\times}10^{-5}$ & $1.09{\times}10^{-4}$ & $7.99{\times}10^{-6}$ & $1.19{\times}10^{-4}$ & $-0.162$ & 39.65 \\
35 & $1.56{\times}10^{-5}$ & $2.01{\times}10^{-5}$ & $1.09{\times}10^{-4}$ & $7.99{\times}10^{-6}$ & $1.19{\times}10^{-4}$ & $-0.161$ & 39.65 \\
\hline\hline
\end{tabular*}
\end{table}

\begin{table}[!htbp]
\centering
\caption{
Influence of maximum values of harmonic mode index $n_{\rm max}$
on the relative uncertainties of intrinsic parameter  in FIM analysis for the case of highly eccentricity $e_{\rm LSO}=0.8$ is listed. The table outlines the relative parameter uncertainties, the correlation coefficient
between $e_{\rm LSO}$ and $\alpha_{\rm DM}$ and the corresponding SNR.
The varied parameters are $\ln M$, $\ln\eta$, $\chi$, $e_{\rm LSO}$, and $\alpha_{\rm DM}$. The other source parameter injected are keeping same with Fig.~\ref{fig:fim-mc-noise}.
}\label{tab:harmonic_truncation_eLSO08}
\scriptsize
\renewcommand{\arraystretch}{1.25}
\setlength{\tabcolsep}{5.5pt}
\begin{tabular*}{\textwidth}{@{\extracolsep{\fill}} cccccccc}
\hline\hline
$n_{\max}$ & $\sigma_M/M$ & $\sigma_\eta/\eta$ & $\sigma_\chi/\chi$ & $\sigma_{e_{\rm LSO}}/e_{\rm LSO}$ & $\sigma_{\alpha_{\rm DM}}/\alpha_{\rm DM}$ & $r_{e\alpha}$ & SNR \\
\hline
1  & $5.03{\times}10^{-3}$ & $1.79{\times}10^{-2}$ & $9.23{\times}10^{-3}$ & $8.29{\times}10^{-4}$ & $8.69{\times}10^{-3}$ & $-0.599$ & 1.04 \\
2  & $3.22{\times}10^{-3}$ & $1.27{\times}10^{-2}$ & $7.31{\times}10^{-3}$ & $5.32{\times}10^{-4}$ & $6.11{\times}10^{-3}$ & $-0.747$ & 1.10 \\
3  & $2.27{\times}10^{-3}$ & $8.11{\times}10^{-3}$ & $6.65{\times}10^{-3}$ & $3.82{\times}10^{-4}$ & $3.96{\times}10^{-3}$ & $-0.655$ & 1.20 \\
4  & $1.33{\times}10^{-3}$ & $3.50{\times}10^{-3}$ & $4.69{\times}10^{-3}$ & $2.30{\times}10^{-4}$ & $1.86{\times}10^{-3}$ & $-0.530$ & 1.81 \\
5  & $8.02{\times}10^{-4}$ & $2.14{\times}10^{-3}$ & $2.78{\times}10^{-3}$ & $1.38{\times}10^{-4}$ & $1.14{\times}10^{-3}$ & $-0.549$ & 2.87 \\
6  & $5.25{\times}10^{-4}$ & $1.23{\times}10^{-3}$ & $1.97{\times}10^{-3}$ & $9.15{\times}10^{-5}$ & $6.82{\times}10^{-4}$ & $-0.463$ & 4.23 \\
10 & $2.60{\times}10^{-4}$ & $4.48{\times}10^{-4}$ & $1.02{\times}10^{-3}$ & $4.56{\times}10^{-5}$ & $2.65{\times}10^{-4}$ & $-0.407$ & 11.07 \\
15 & $1.49{\times}10^{-4}$ & $2.13{\times}10^{-4}$ & $5.89{\times}10^{-4}$ & $2.62{\times}10^{-5}$ & $1.26{\times}10^{-4}$ & $-0.400$ & 20.18 \\
21 & $1.20{\times}10^{-4}$ & $1.51{\times}10^{-4}$ & $4.91{\times}10^{-4}$ & $2.13{\times}10^{-5}$ & $8.58{\times}10^{-5}$ & $-0.331$ & 24.62 \\
25 & $1.18{\times}10^{-4}$ & $1.41{\times}10^{-4}$ & $4.83{\times}10^{-4}$ & $2.08{\times}10^{-5}$ & $7.96{\times}10^{-5}$ & $-0.329$ & 25.12 \\
30 & $1.17{\times}10^{-4}$ & $1.38{\times}10^{-4}$ & $4.83{\times}10^{-4}$ & $2.07{\times}10^{-5}$ & $7.63{\times}10^{-5}$ & $-0.320$ & 25.19 \\
35 & $1.18{\times}10^{-4}$ & $1.33{\times}10^{-4}$ & $4.84{\times}10^{-4}$ & $2.08{\times}10^{-5}$ & $7.37{\times}10^{-5}$ & $-0.335$ & 25.20 \\
\hline\hline
\end{tabular*}
\end{table}